\documentclass[aps,prd,amsmath,amssymb,preprint,groupedaddress]{revtex4-1}
\usepackage{graphicx}
\usepackage{mathrsfs}
\usepackage{epsfig}
\def\comment#1{}
\def\bn{\bigskip\noindent}

\begin{document}

\title[Fast Evolution and Waveform Generator for EMRIs in Equatorial-Circular Orbits]{Fast Evolution and Waveform Generator for Extreme-Mass-Ratio Inspirals in Equatorial-Circular Orbits}

\author{Wen-Biao Han}
\email{wbhan@shao.ac.cn}
\affiliation{Shanghai Astronomical Observatory, Chinese Academy of Sciences \\
80 Nandan Road, Shanghai, China P.R., 200030}

\begin{abstract}
In this paper we discuss the development of a fast and accurate waveform model for the quasi-circular orbital evolution of extreme-mass-ratio-inspirals (EMRIs). This model simply employs the data of a few numerical Teukoulsky-based energy fluxes and waveforms to fit out a set of polynomials for the entire fluxes and waveforms. These obtained polynomials are accurate enough in the entire evolution domain, and much more accurate than the resummation post-Newtonian (PN) energy fluxes and waveforms, especially when the spin of a black hole becomes large. The dynamical equation  we adopted for orbital revolution is the effective-one-body (EOB) formalism. Because of the simplified expressions, the efficiency of calculating the orbital evolution with our polynomials is also better than the traditional method which uses the resummed PN analytical fluxes. Our model should be useful in calculation of waveform templates of EMRIs for the gravitational wave detectors such as the evolved Laser Interferometer Space Antenna (eLISA).
\end{abstract}

%
%
%
%
%
\maketitle

\section{Introduction\label{intro}}

An extreme-mass-ratio inspiral (EMRI) arises following the capture of a compact star with stellar mass (white dwarf, neutron star or black hole) by a supermassive black hole. The orbital radius of such EMRI is about or less than $O(10^1)$ Schwarzschild radius of the supermassive black hole. Because of gravitational radiation, the orbit of the small body shrinks toward the central black hole in a long time scale. EMRIs are very important for revealing the properties of supermassive black holes since people may observe the gravitational signals with many wave cycles from the region near the horizon of the black hole. EMRIs are potential sources for eLISA (evolved Laser Interferometer Space Antenna), a space gravitational wave observatory supported by the European Space Agency now \cite{elisa}. A pathfinder has been launched in 2015 to pave the way for eLISA.

Due to the match-filter technology employed in gravitational wave detection, people must have a huge amount of theoretical waveform templates with enough accuracy in a very large parameter space. Up to now, there are usually three methods to compute the theoretical waveforms: the first one is the post-Newtonian (PN) approximation, the second one is numerical relativity, and the last one is black hole perturbation theory. As analyzed in a great deal of literatures (for example see \cite{Hinderer}), for EMRIs, the PN expansion would lose accuracy greatly in such a highly relativistic region, and even the factorized-resummed PN waveforms in an effective-one-body (EOB--an analytical approach which aims at providing an accurate description of the motion and radiation of coalescing binary black holes with arbitrary mass ratio, see review \cite{damour12} for details) frame also do not have a good performances for spinning black holes (see for example Ref. \cite{Panyi11} and references insides); Numerical relativity still can not handle the binary black hole systems with extreme mass ratio. An recent paper documents the simulation of a case of  mass ratio 1:100 (without spin) with only two orbits' evolution \cite{Lousto11}.

Therefore, for such small mass-ratio binaries, the black hole perturbation theory is a good tool to study EMRIs (mass-ratio $\sim 10^{-7}-10^{-4}$) \cite{Poisson1,Poisson2,Shibata,Hughes1,Hughes2,Glampedakis1,Hughes3,Hughes4,Fujita3,Yanbei12,Hughes14} and even IMRIs (mass-ratio $\sim 10^{-3} - 10^{-2}$) \cite{Nagar07,Bernuzzi10,Bernuzzi11a,Bernuzzi11b,Han11,Barausse12}. It means that the small body can be treated as a perturbation of the background field of the central supermassive black hole. The black hole perturbation theory was built by Regge, Wheeler and Zerilli in Schwarzschild spacetime \cite{R-W,Zerilli} and by Teukolsky in Kerr background \cite{Teukolsky1,Teukolsky2}.

Usually, the eLISA observes wave-signals over a span of several years. This means that one needs to model the waveforms of EMRIs over a length of several years. Though the cost of calculation of the Teukolsky equation is much less than numerical relativity, it is still unaffordable for evolving $O(10^5-10^6)$ orbits in various parameters. Therefore, researchers are challenged about how to greatly reduce the CPU time of the simulation of EMRI waveform, while at the same time maintaining enough accuracy. For the quasi-circular orbits, Yunes et al used Teukolsky-based energy fluxes to fit higher order pN fluxes \cite{Yunes10,Yunes11}. Using the self-force data of Schwarzschild black hole, Lackeos and Burko fitted polynomials mixed PN expressions to do the orbital evolution of IMRIs, but they still numerically solved the Teukolsky equation to get the waveforms \cite{Lackeos12}. Fujita gave out the expressions of gravitational radiation up to 14th PN order by computing the Teukolsky equation analytically \cite{Fujita12}. However, in this paper, we completely abandon using the analytical PN expansions. We use the Teukolsky-based numerical data to directly fit out a set of polynomials for energy fluxes and waveforms. This method is very simple and efficient, and the fitted polynomials can give very highly accurate energy-fluxes and waveforms.  In principle, these polynomials are also a kind of PN expansions, but all the coefficients of such ``PN expansions'' are obtained numerically from the fitting of the Teukolsky-based data.

In the next section, we introduce our EOB-Teukolsky codes (ET codes) shortly. The details of our fitting polynomial method are presented in the section 3. In section 4, results and comparisons are shown. Finally, conclusions and remarks are given in section 5. Throughout the paper, we use units $G=c=1$ and the metric signature $(-,+,+,+)$. Distance
and time are measured by the central black-hole mass $M$.

\section{EOB-Teukolsky frequency-domain codes}
Our ET codes include two main parts: one is the EOB dynamics driver, the other is the Teukolsky equation solver. The EOB  part gives the orbital parameters to the Teukolsky equation, and then the later one produces waveforms and energy-fluxes. Next the Teukolsky-based energy-fluxes source the EOB dynamics to drive the orbital evolution of the small body around the supermassive black hole \cite{Han11,Han14}. A detailed introduction of the EOB dynamics will be presented in the section 4.

The Teukolsky equation in ET codes can be solved in frequency-domain for inspiralling phase and in time-domain for plunge and merge states. In the present work, we focus on only inspiralling process, we just need the frequency-domain Teukolsky calculation. We employ a semi-analytical method to solve the frequency-domain equation which was developed in \cite{MST,Sasaki2,Fujita1,Fujita2} to replace the previous numerical integration method \cite{Han10}.

After decomposing the Weyl curvature
(complex) scalar $\psi_4$ in a Fourier series \cite{Teukolsky1},
\begin{align}
\psi_4=\rho^4\int^{+\infty}_{-\infty}{d\omega\sum_{lm}{R_{lm\omega}(r)_{~-2}S^{a\omega}_{lm}(\theta)e^{im\phi}e^{-i\omega
t}}},
\end{align}
where $\rho=-1/(r-ia\cos{\theta})$, the Teukolsky equation is divided into two parts. One is the radial master equation
\begin{align}
\Delta^2\frac{d}{dr}\left(\frac{1}{\Delta}\frac{d
R_{lm\omega}}{dr}\right)-V(r)R_{lm\omega}=-\mathscr{T}_{lm\omega}(r),\label{Teukolskyr}
\end{align}
where $\mathscr{T}_{lm\omega}(r)$ is the source term which is connected with the energy-momentum tensor of the test particle around a black hole, and the potential is
\begin{align}
V(r)=-\frac{K^2+4i(r-M)K}{\Delta}+8i\omega r+\lambda,
\end{align}
where $K=(r^2+a^2)\omega-ma, ~\lambda=E_{lm}+a^2\omega^2-2a m w-2$. The other is the angular equation
\begin{align}
&\frac{1}{\sin{\theta}}\frac{d}{d\theta}\left(\sin{\theta
\frac{d_{~-2}S^{a\omega}_{lm}}{d\theta}}\right) + \nonumber \\
&\left[(a\omega)^2\cos^2{\theta}+4a\omega\cos{\theta}-\left(\frac{m^2-4m\cos\theta+4}{\sin^2\theta}\right)+E_{lm}\right]
~_{-2}S^{a\omega}_{lm}=0 ,\label{spinweightedspheroidal}
\end{align}
where $_{-2}S^{a\omega}_{lm}(\theta)$ is the spin-weighted angular function.

If $\mathscr{T}_{lm\omega}(r) = 0$, Eq. (\ref{Teukolskyr}) becomes a homogeneous equation, then can be solved quickly and accurately by a semi-analytical
numerical method developed by Fujita and
Tagoshi \cite{Fujita1,Fujita2}. The homogeneous
solutions of Teukolsky radial equation are expressed in terms of two kinds
of series of special functions. The first one consists of series of
hypergeometric functions and is convergent at the horizon
\begin{align}
R_{lm\omega}^{\text{H}}=e^{i\epsilon\kappa
x}(-x)^{-s-i(\epsilon+\tau)/2}(1-x)^{i(\epsilon-\tau)/2}p_{\text{in}}(x),
\end{align}
where $p_{\text{in}}$ is expanded in a series of hypergeometric
functions as
\begin{align}
p_{in}(x)=\sum_{n=-\infty}^{\infty}{a_n F(n+\nu+1-i\tau,
-n-\nu-i\tau;1-s-i\epsilon-i\tau;x)},\label{hypergeo}
\end{align}
and $x=\omega(r_+-r)/\epsilon\kappa$,
$\epsilon=2M\omega$,$\kappa=\sqrt{1-a^2}$,$\tau=(\epsilon-m
a)/\kappa$. The hypergeometric function $F(\alpha,\beta;\gamma;x)$
can be found in mathematic handbooks.

The second one consists of series of Coulomb wave functions which is
convergent at infinity. The homogeneous solution of Teukolsky
equation is
\begin{align}
R_C=z^{-1-s}(1-\epsilon\kappa/z)^{-s-i(\epsilon+\tau)/2}f_\nu(z),
\end{align}
where $f_\nu(z)$ is expressed in a series of Coulamb wave functions
as
\begin{align}
f_\nu(z)=\sum_{-\infty}^{\infty}{(-i)^n\frac{(\nu+1+s-i\epsilon)_n}{(\nu+1-s+i\epsilon)_n}a_nF_{n+\nu}(-is-\epsilon,z)},\label{coulomb}
\end{align}
and $z=\omega(r-r_-),(a)_n=\Gamma(a+n)/\Gamma(a)$, $F_N(\eta,z)$ is
a Coulomb wave function. The outgoing homogeneous solution can be expressed in Coulomb wave functions as,
\begin{align}
R^{\infty}_{lm\omega}=A^{\nu}_- z^{-1-2s}e^{i(z+\epsilon \ln{z})},
\end{align}
where the coefficient $A^{\nu}_-$ is
\begin{align}
A^{\nu}_-=2^{-1-s+i\epsilon}e^{-\pi\epsilon/2}e^{-i\pi(\nu+1-s)/2}\sum^{+\infty}_{-\infty}{(-1)^n \frac{(\nu+1+s-i\epsilon)_n}{(\nu+1-s+i\epsilon)_n}}f_n^\nu.\label{Rout}
\end{align}
Note that Eqs.(\ref{hypergeo}-\ref{Rout}) involve a parameter $\nu$, the
so-called renormalized angular momentum. The key part of this semi-analytical method is to search the renormalized angular momentum numerically (see \cite{Fujita1, Fujita2} for details).

With the $R^{\infty}_{lm\omega}(r)$ and $R^{H}_{lm\omega}(r)$ in hand,
take into the source term $\mathscr{T}_{lm\omega}$ (for an equatorial-circular case, see our previous work \cite{Han10}),
\begin{align} \nonumber
\mathscr{T}_{lm\omega}(r) = \int dt \Delta^2
\big\{&(A_{nn0}+A_{n\bar{m}0}+A_{\bar{m}\bar{m}0})\delta(r-r_0)+ \\
&\partial_r[(A_{n\bar{m}1}+A_{\bar{m}\bar{m}1})\delta(r-r_0)]+\partial_r^2[A_{\bar{m}\bar{m}2}\delta(r-r_0)] \big\}.\label{source}
\end{align}

we have
\comment{\begin{align}\nonumber
Z^{\text{H}}_{lm\omega}&=\frac{1}{2i\omega
B^{\text{in}}_{lm\omega}}\int^{+\infty}_{-\infty}{dt e^{i\omega t-im\phi(t)}\mathcal{I}^H_{lm\omega}(r(t),\theta(t))},\\
Z^{\infty}_{lm\omega}&=\frac{B^{\text{hole}}_{lm\omega}}{2i\omega
B^{\text{in}}_{lm\omega}D^{\infty}_{lm\omega}}\int^{+\infty}_{-\infty}{dt e^{i\omega t-im\phi(t)}\mathcal{I}^\infty_{lm\omega}(r(t),\theta(t))}.\label{Z1}
\end{align}
}
\begin{align}
Z^{H}_{lm} &=\frac{\pi}{i\omega_{m}
B^{\text{in}}_{lm\omega}T_r}\mathcal{I}^{H}(r_0),\label{Zh} \\
Z^{\infty}_{lm} &=\frac{\pi B^{\text{hole}}}{i\omega_{m}
B^{\text{in}}_{lm\omega}D^{\infty}_{lm\omega}T_r} \mathcal{I}^{\infty}(r_0),\label{Z8}
\end{align}

where
\begin{align}
\mathcal{I}^{H,\infty}_{lm\omega}=\left[R^{H,\infty}_{lm\omega}(A_{nn0}+A_{\bar{m}n0}+A_{\bar{m}\bar{m}0})-\frac{dR^{H,\infty}_{lm\omega}}{dt}(A_{\bar{m}n1}+A_{\bar{m}\bar{m}1})+\frac{d^2R^{H,\infty}_{lm\omega}}{dt^2}A_{\bar{m}\bar{m}2}\right]_{r_0,\theta=\pi/2}
\end{align}

All the quantities ($A_{nn0}, A_{\bar{m}n0}$, etc.) shown in the above equations were given explicitly in \cite{Hughes1},  and the harmonic
frequency is $\omega_{m}=m\Omega_\phi $. The calculation of coefficients $A_{nn0}, A_{\bar{m}n0}, \cdots $ involves the solution of the angular equation (\ref{spinweightedspheroidal}). There are several routes to calculate the spin weighted spheroidal function $_{~-2}S^{a\omega}_{lm}$. In this paper, we adopt the method described in \cite{Hughes1}.

\comment{
Then $Z^{H,\infty}_{lm\omega}$ are decomposed as
\begin{align}\nonumber
Z^{\text{H}}_{lm\omega}&=Z^{\text{H}}_{lm}\delta(\omega-\omega_{m}),\\
Z^{\infty}_{lm\omega}&=Z^{\infty}_{lm}\delta(\omega-\omega_{m}).\label{Z2}
\end{align}
}
Then the amplitudes $Z^{H,\infty}_{lm}$ fully determine the
fluxes of gravitational radiations to infinity and horizon,
\begin{align}
\dot{E}^{\infty,\text{H}}&=\sum_{lm}{\frac{|Z^{\mathrm{H},\infty}_{lm}|^2}{4\pi\omega^2_{m}}},\label{energyflux} \\
\dot{L_z}^{\infty,\text{H}}&=\sum_{lm}{\frac{m|Z^{\mathrm{H},\infty}_{lm}|^2}{4\pi\omega^3_{m}}},\label{angularflux}
\end{align}
where the overdot stands for ${\rm d}/{\rm d} t$, $E$ and $L_z$ mean energy and angular momentum respectively. The notations $\infty$ and H on $E$ and $L_z$ mean the fluxes to infinity and the horizon respectively. The gravitational waveform can be expressed as:
\begin{align}
h_+-ih_\times=\frac{2}{r}\sum_{lm}{\frac{Z^{\mathrm{H}}_{lm\omega}}{\omega^2_{m}}S^{a\omega_{m}}_{lm}(\theta)e^{-i\omega_{m}t+im\phi}}. \label{waveform}
\end{align}

\section{Fitting polynomial from Teukolsky-based energy fluxes and waveforms}
As we mentioned, the CPU expense for simulating the EMRI evolution by numerical solving the Teukolsky equation is quite huge. Adopting PN fluxes to do evolution is fast but will lost the accuracy when the small body approaching it's innermost stable circular orbit (ISCO) around the central black hole. In this section, we use the Teukolsky-based flux data of a few points to fit out a set of polynomials for replacing the PN fluxes or original Teukolsky-based fluxes. The idea is very simple: we select a few points between the initial radius and ISCO, the calculate the Teukolsky-based fluxes and waveforms at these points, and then use these data to fit out a set of polynomials. The fitted polynomials are used as fluxes and waveforms, they are functions of the radius $r$ to the black hole.

We choose 7, 9, 11 and 13 points (Chebshev nodes) to fit out the 6th, 8th, 10th and 12th order polynomials and compare the results with factorized PN ones (of course, one also can use more points to fit out these polynomials). For completeness, we need four set of polynomials for the orbital evolution and waveform extraction. Two of them are for fluxes down to the infinity and horizon:
\begin{align} \label{Epoly}
\dot{E}^{\infty}=\sum_{i=0}^{n}a_i /r^i,\quad \dot{E}^{\text{H}}=\sum_{i=0}^{n}b_i /r^i \,,
\end{align}
where $n$ is the order of polynomials. For waveforms, from Eq. (\ref{waveform}), the term can be fitted by polynomials is
\begin{align*}
 H_{lm}  \equiv \frac{Z^{\mathrm{H}}_{lm\omega}}{\omega^2_{m}}S^{a\omega_{m}}_{lm}(\theta) .
\end{align*}
Considering this term is a complex function, we need to use
 $2 \times (n+1)$ fitting polynomials for each $(l,m)$ mode:
\begin{align}\label{hpoly}
\text{Re}[H_{lm}]=\sum_{i=0}^{n} R^{i}_{lm} /r^i , \quad \text{Im}[H_{lm}]=\sum_{i=0}^{n} I^{i}_{lm} /r^i,
\end{align}
where $R^{i}_{lm}$ and $I^{i}_{lm}$ are the polynomial coefficients.

For reducing interpolation errors, we use the Chebshev nodes to produce the interpolating points in our model:
\begin{align}
r_i = \frac{r_b+r_a}{2}+\frac{r_b-r_a}{2}\cos{\frac{2i+1}{2n'+1}\pi} \,,
\end{align}
where $r_a$, $r_b$ are the boundaries of the calculation area and $n'$ is the total number of nodes, and $i$ goes from 0 to $n' -1$.

\begin{figure}
\begin{center}
\includegraphics[height=2.0in]{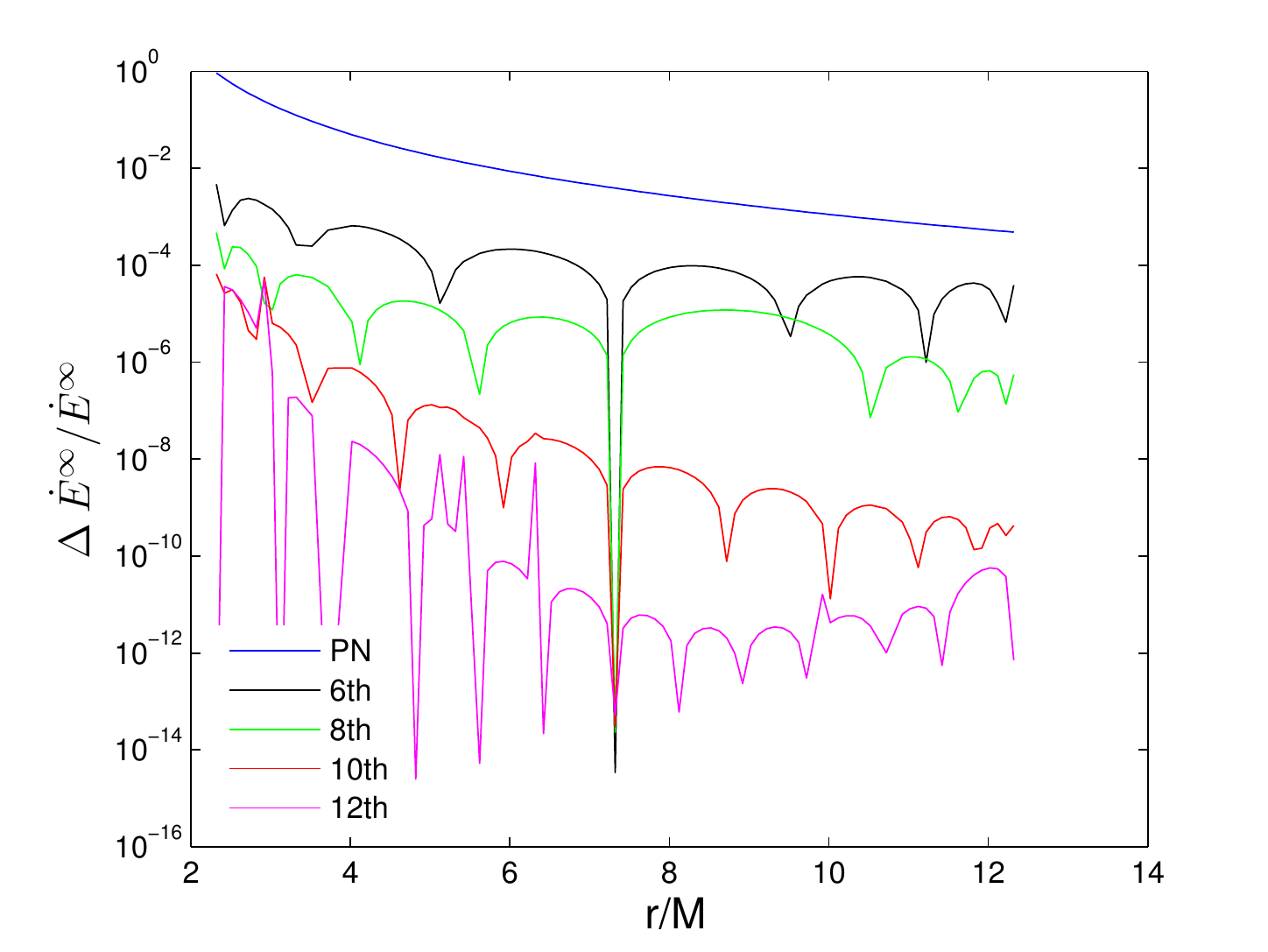}
\includegraphics[height=2.0in]{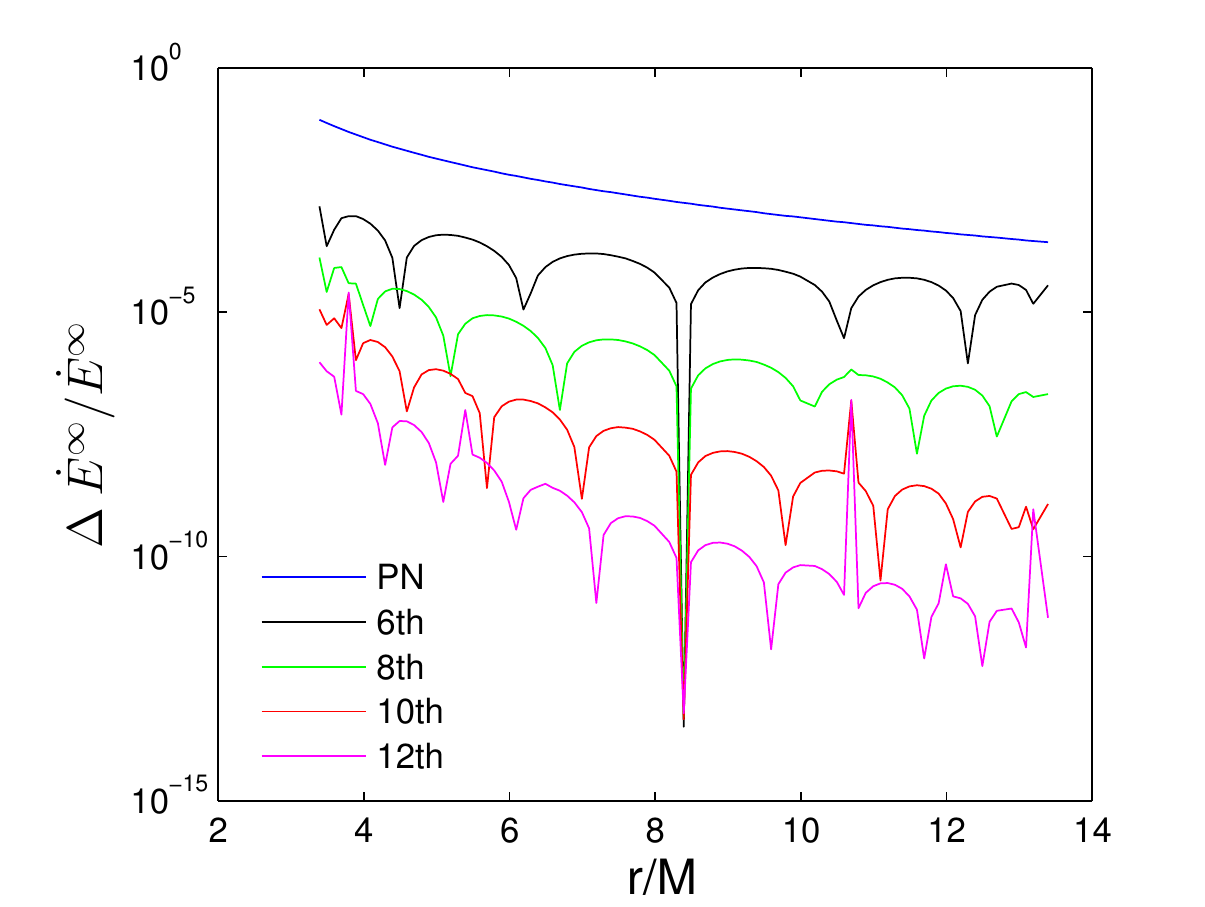}
\includegraphics[height=2.0in]{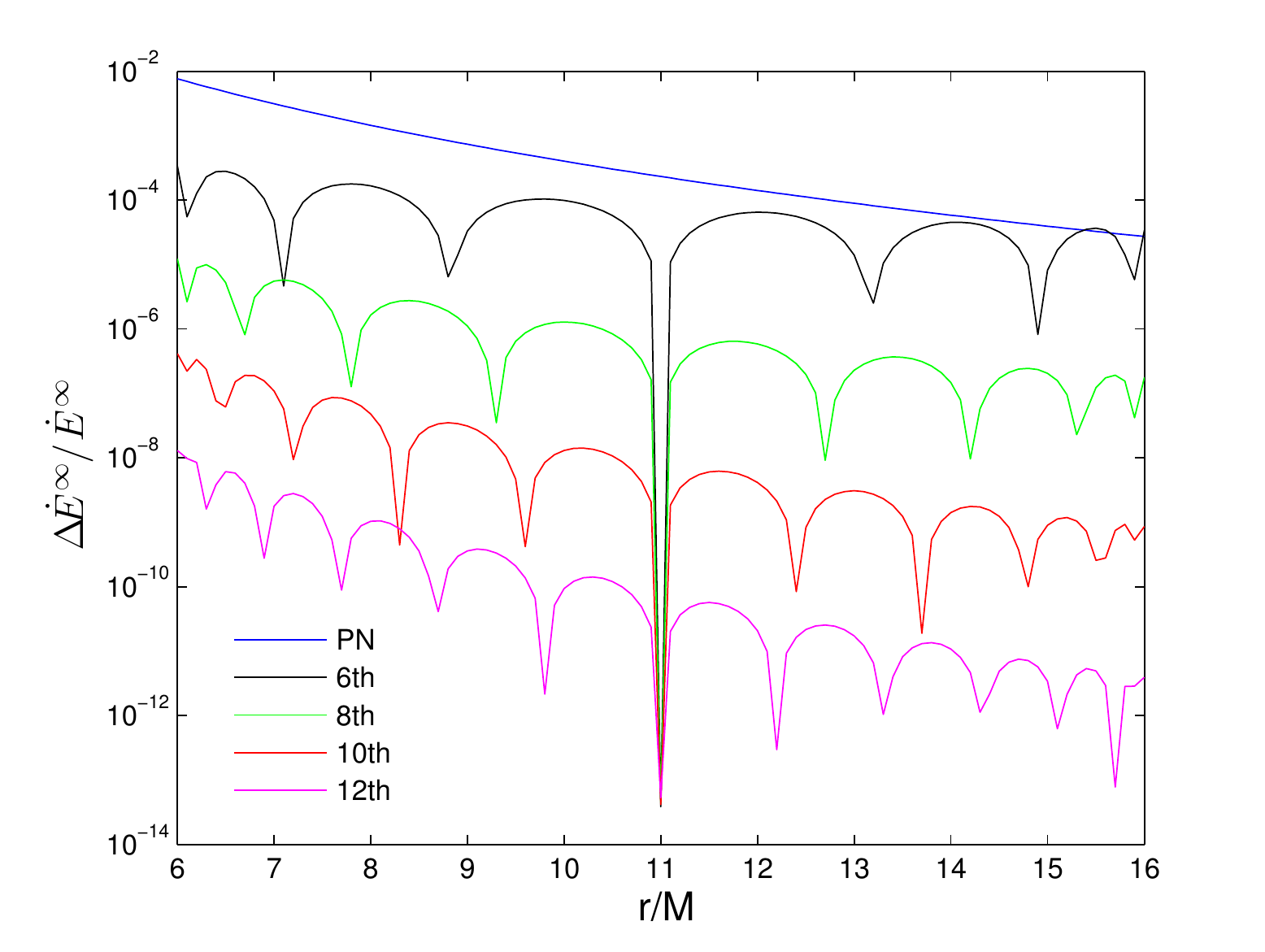}
\includegraphics[height=2.0in]{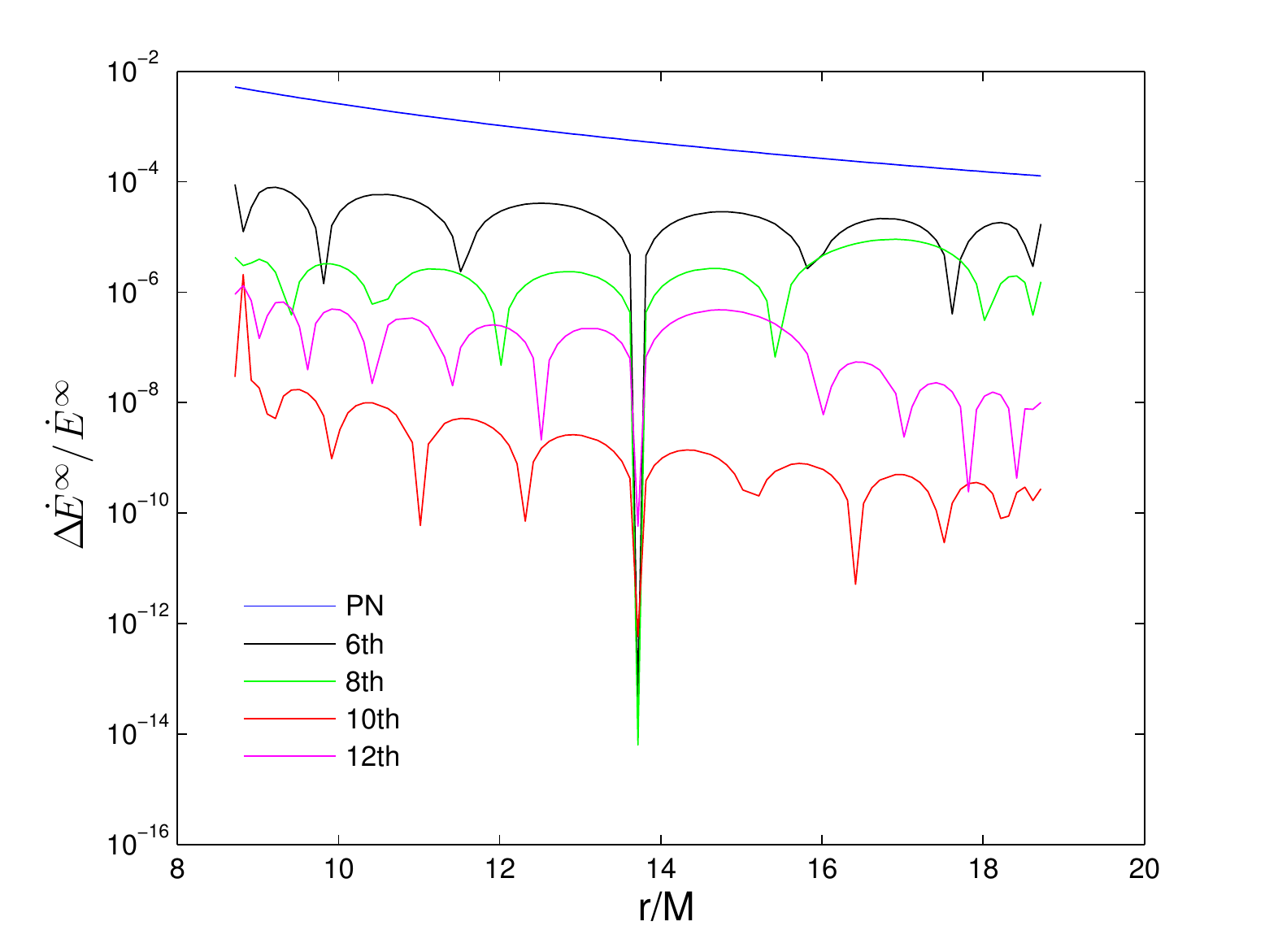}
\caption{Comparing the 6th, 8th, 10th, 12th polynomials and factorized PN energy fluxes to infinity for $a=0.9, ~0.7, ~0, ~-0.9$ (from left to right, top to bottom). $\Delta \dot{E}^\infty$ is difference between the fitted polynomials or PN energy fluxes with the accurate numerical Teukolsky date $\dot{E}^\infty$.}  \label{E8fit}
\end{center}
\end{figure}

\begin{figure}
\begin{center}
\includegraphics[height=2.0in]{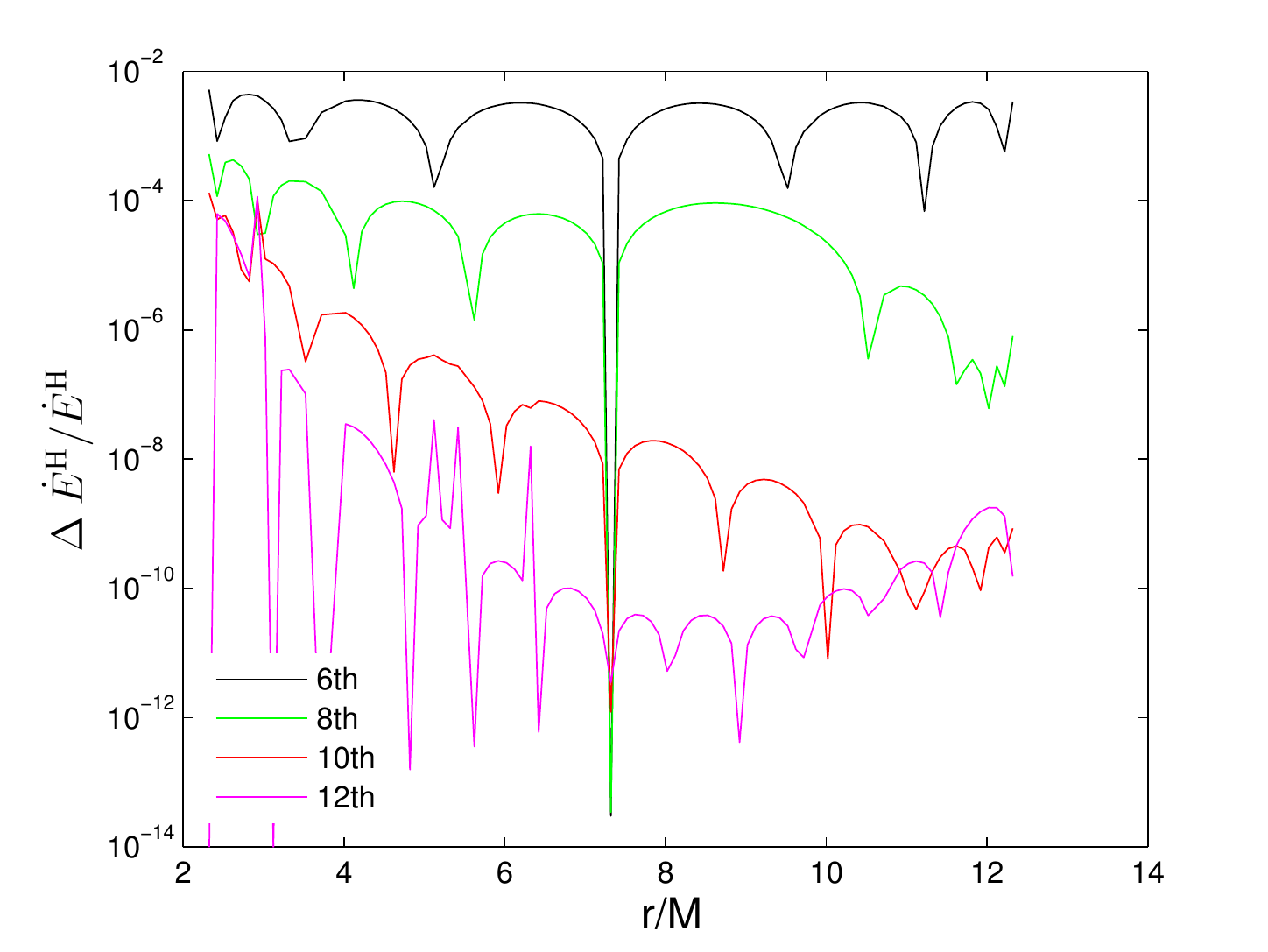}
\includegraphics[height=2.0in]{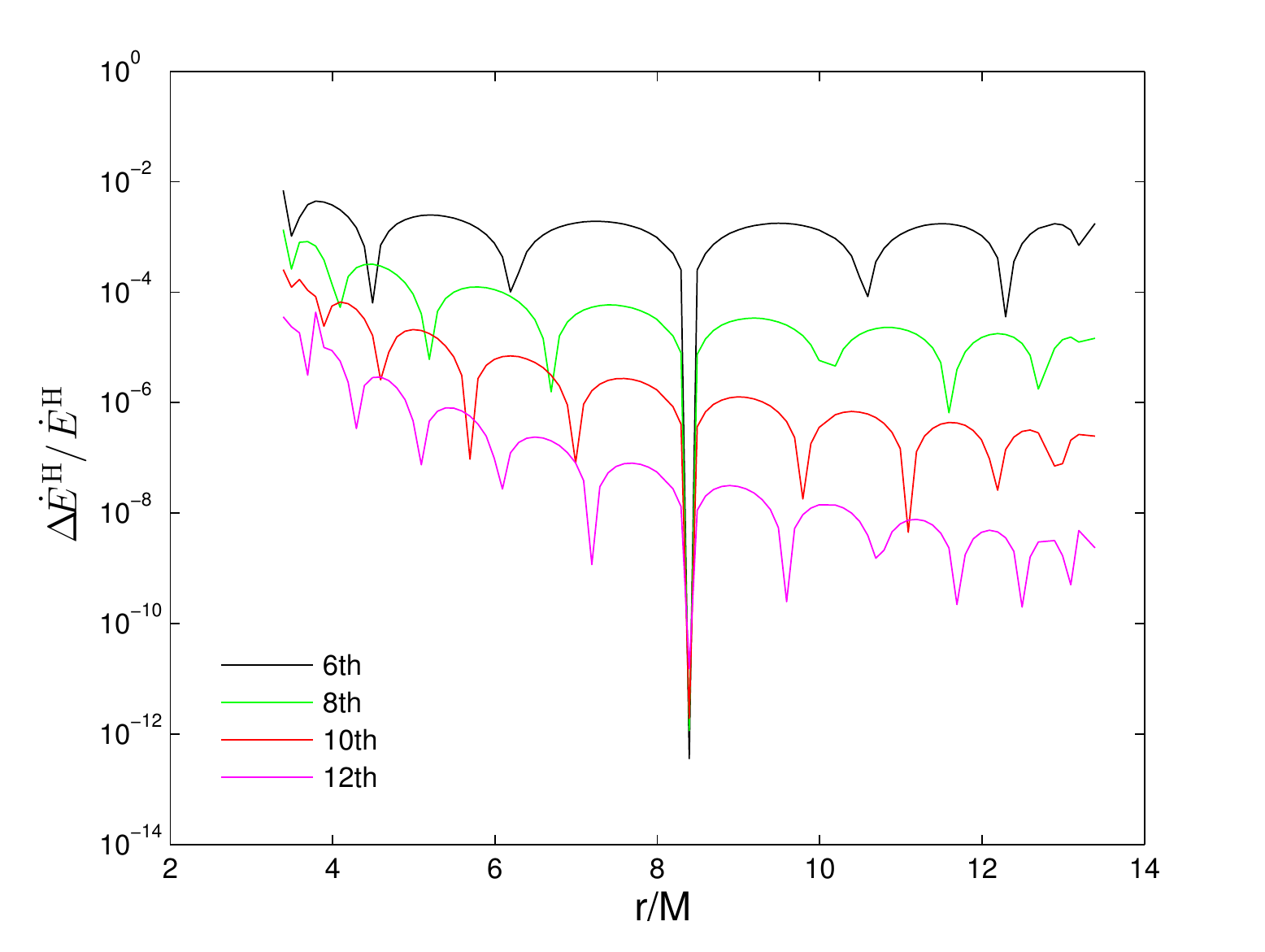}
\includegraphics[height=2.0in]{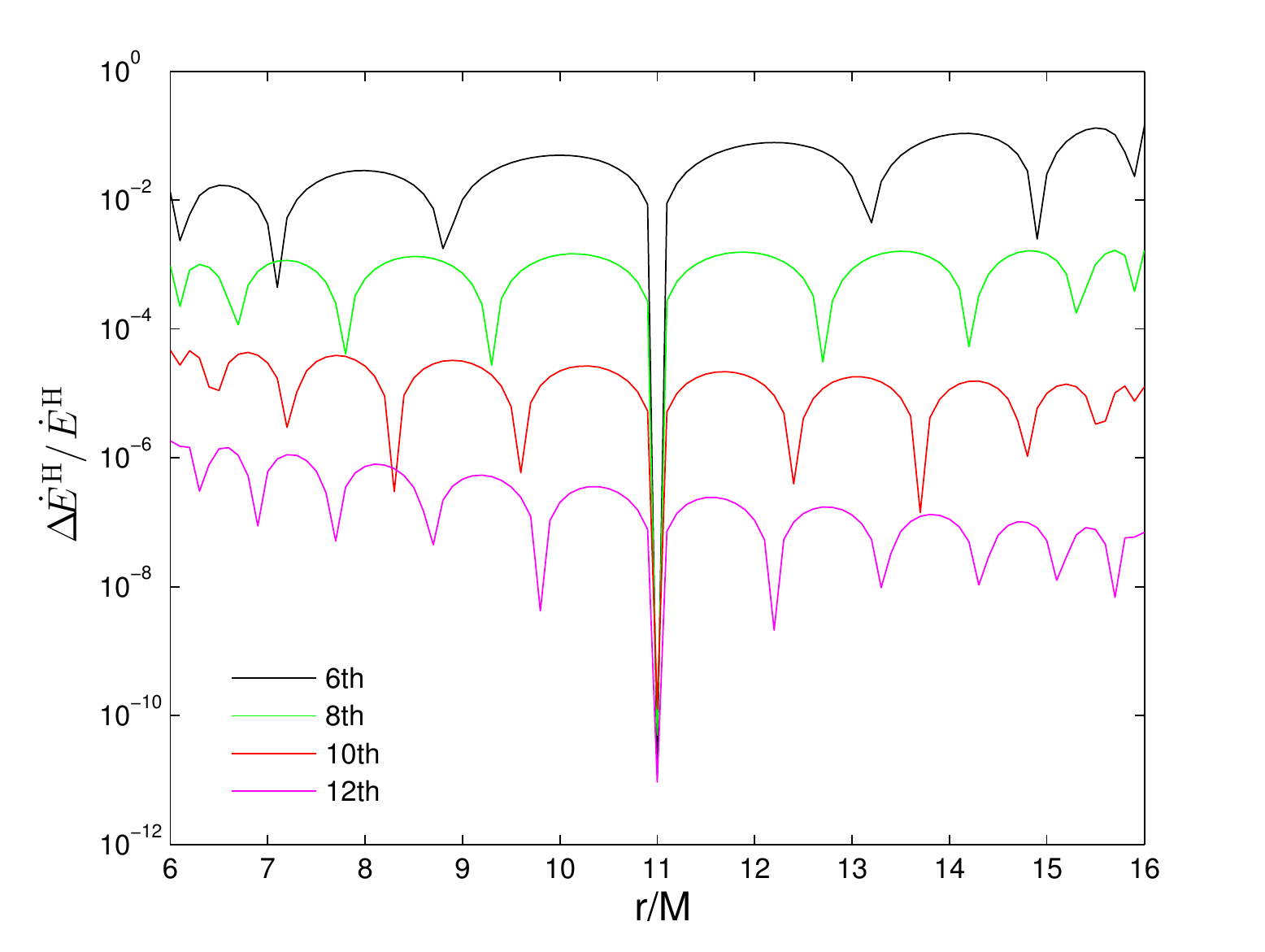}
\includegraphics[height=2.0in]{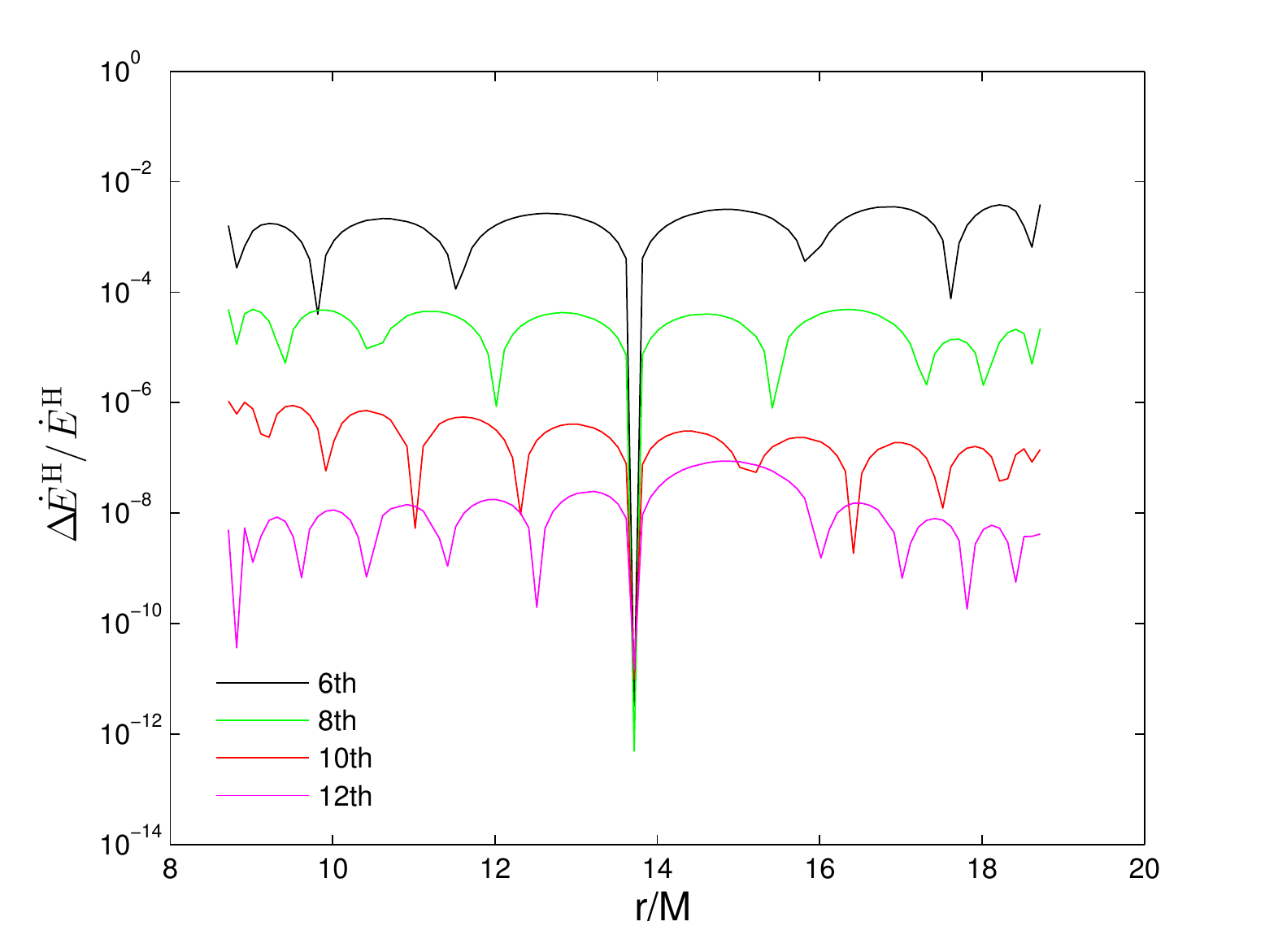}
\caption{Comparing the 6th, 8th, 10th, 12th polynomials fitted from energy fluxes to horizon for $a=0.9, ~0.7, ~0, ~-0.9$ (from left to right, top to bottom). $\Delta \dot{E}_\text{H}$ is difference between the fitted polynomial energy fluxes with the accurate numerical Teukolsky data $\dot{E}_\text{H}$.}  \label{Ehfit}
\end{center}
\end{figure}

In Fig. \ref{E8fit} and Fig. \ref{Ehfit}, we compare the fitted polynomials of different orders with the factorized-resummation PN fluxes. We can find that the 6th polynomials do not give a good performance.  And the factorized-resummation PN fluxes which are used in the EOB model perform worst. Both the 10th and 12th order polynomial can give very good fit to the numerical Teukolsky-based fluxes. It is assumed that LISA will observe gravitational waves of EMRIs at the typical frequency $\sim 10^2$ Hz and the total wave cycle is about $N  \sim10^5$ for 1 yr. Thus, the relative error of energy luminosity required to establish the accuracy for the cycle $\Delta N \le 1$ must be $\leq 10^{-5}$ in circular orbit cases \cite{cutler93}. Therefore, the 6th polynomials cannot satisfy the requirement of accuracy. The 8th one is at the edge of this requirement. The 10th polynomials can meet the requirement well. Considering the simplification and saving CPU time, we decide to use the 10th polynomial to fit energy fluxes and waveform in this paper. We list the 10th polynomial coefficients of energy fluxes in Tab. \ref{E8a9} and Tab. \ref{Eha9} which are used in Fig. \ref{E8fit} and Fig. \ref{Ehfit}. The polynomial coefficients for calculating (2, 2)-mode waveform displayed in Fig. \ref{waveforms} are listed in Tab. \ref{RH22} and \ref{IH22}.

\begin{table}
  \caption{polynomial parameters for infinity fluxes.}\label{E8a9}
  \resizebox{\textwidth}{!}{%
  \begin{tabular}{l|ccccccccccc}
  \hline\hline
  \quad & $a_0$ & $a_1$ & $a_2$ & $a_3$ & $a_4$ & $a_5$ & $a_6$ & $a_7$ & $a_8$ & $a_9$ & $a_{10}$ \\
  \hline
  $a=0.9$ & 1.51e-7 & -1.22e-5 & 4.40e-4 & -9.43e-3 & 1.36e-1 & 4.90e0 & -5.27e0 & -1.05e1 & 2.98e1 & -4.34e1 & 3.99e1 \\
  \hline
  $a=0.7$ & 6.61e-7 & -5.98e-5 & 2.41e-3 & -5.69e-2 & 8.76e-1 & -2.90e0 & 5.22e1 & -2.75e2 & 8.81e2 & -1.64e3 &1.44e3 \\
  \hline
  $a=0.0$ & 1.46e-6 & -1.72e-4 & 9.08e-3 & -2.83e-1 & 5.76e0 & -7.39e1 & 7.63e2 & -5.02e3 & 2.21e4 & -5.77e4 & 7.19e4 \\
  \hline
  $a=-0.9$ & 1.26e-6 & -1.85e-4 & 1.22e-2 & -4.78e-1 & 1.22e1  & -2.09e2 & 2.63e3 & -2.21e4 & 1.25e5 & -4.20e5 & 6.78e5 \\
  \hline\hline
  \end{tabular}
  }
\end{table}

\begin{table}
  \centering
  \caption{polynomial parameters for horizon fluxes.}\label{Eha9}
  \resizebox{\textwidth}{!}{%
  \begin{tabular}{l|ccccccccccc}
  \hline\hline
  \quad & $b_0$ & $b_1$ & $b_2$ & $b_3$ & $b_4$ & $b_5$ & $b_6$ & $b_7$ & $b_8$ & $b_9$ & $b_{10}$ \\
  \hline
  $a=0.9$ & -3.44e-9 & 2.93e-7 & -1.09e-5 & 2.25e-4 & -2.66e-3 & 1.34e-2 & 1.07e-1 & -3.19e0 & 8.84e0 & -8.33e0 & 2.37e0 \\
  \hline
  $a=0.7$ & 8.88e-8 & -7.97e-6 & 3.17e-4 & -7.39e-3 & 1.11e-1 & -1.14e0 & 8.00e0 & -3.89e1 & 1.17e2 & -2.04e2 & 1.58e2 \\
  \hline
  $a=0.0$ & 4.84e-7 & -5.70e-5 & 3.00e-3 & -9.28e-2 & 1.88e0 & -2.59e1 & 2.48e2 & -1.62e3 & 6.97e3 & -1.79e4 & 2.11e4 \\
  \hline
  $a=-0.9$ & 5.30e-7 & -7.76e-5 & 5.11e-3 & -1.99e-1 & 5.08e0 & -8.90e1 & 1.08e3 & -9.10e3 & 5.05e4 & -1.69e5 & 2.63e5 \\
  \hline\hline
  \end{tabular}
  }
\end{table}

\begin{table}
  \centering
  \caption{polynomial coefficients for waveform (2,2) mode: real part.}\label{RH22}
  \resizebox{\textwidth}{!}{%
  \begin{tabular}{l|ccccccccccc}
  \hline\hline
  \quad & $R^0_{22}$ & $R^1_{22}$ & $R^2_{22}$ & $R^3_{22}$ & $R^4_{22}$ & $R^5_{22}$ & $R^6_{22}$ & $R^7_{22}$ & $R^8_{22}$ & $R^9_{22}$ & $R^{10}_{22}$ \\
  \hline
  $a=0.9$ & 4.00e-5 & 4.96e-1 & -9.20e-1 & -3.43e-1 & -6.23e0 & 1.73e2 & -1.02e3 & 3.20e3 & -6.14e3 & 6.78e3 & -3.27e3 \\
  \hline
  $a=0.7$ & 2.88e-5 & 4.96e-1 & -9.20e-1 & 1.03e0 & -1.98e1 & 2.75e2 & -1.57e3 & 5.33e3 & -1.18e4 & 1.54e4 & -9.09e3 \\
  \hline
  $a=0.0$ & 1.78e-5 & 4.97e-1 & -8.67e-1 & 5.44e0 & -6.22e1 & 6.73e2 & -4.23e3 & 1.89e4 & -5.58e4 & 9.83e4 & -7.12e4 \\
  \hline
  $a=-0.9$ & 1.38e-5 & 4.97e-1 & -7.79e-1 & 1.17e1 & -1.17e2 & 1.34e3 & -9.37e3 & 4.84e4 & -1.46e5 & 2.09e5 & 1.13e5 \\
  \hline\hline
  \end{tabular}
  }
\end{table}

\begin{table}
  \centering
  \caption{polynomial coefficients for waveform (2,2) mode: imaginary part.}\label{IH22}
  \resizebox{\textwidth}{!}{%
  \begin{tabular}{l|ccccccccccc}
  \hline
  \quad & $I^0_{22}$ & $I^1_{22}$ & $I^2_{22}$ & $I^3_{22}$ & $I^4_{22}$ & $I^5_{22}$ & $I^6_{22}$ & $I^7_{22}$ & $I^8_{22}$ & $I^9_{22}$ & $I^{10}_{22}$ \\
  \hline
  $a=0.9$ & -8.83e-5 & 1.20e-2 & -1.72e0 & 2.85e0 & 4.48e1 & -3.08e2 & 1.16e3 & -2.89e3 & 4.82e3 & -5.07e3 & 2.49e3 \\
  \hline
  $a=0.7$ & -6.68e-5 & 1.03e-2 & -1.66e0 & 1.52e0 & 5.76e1 & -4.20e2 & 1.91e3 & -6.10e3 & 1.36e4 & -1.88e4 & 1.18e4 \\
  \hline
  $a=0.0$ & -3.52e-5 & 7.18e-3 & -1.52e-1 & -2.30e0 & 1.02e2 & -9.80e2 & 6.81e3 & -3.42e4 & 1.19e5 & -2.53e5 & 2.54e5 \\
  \hline
  $a=-0.9$ & -2.06e-5 & 5.32e-3 & -1.41e0 & -6.21e0 & 1.59e2 & -2.02e3 & 1.86e4 & -1.24e5 & 5.72e5 & -1.62e6 & 2.20e6 \\
  \hline\hline
  \end{tabular}
  }
\end{table}

\begin{figure}
\begin{center}
\includegraphics[height=2.0in]{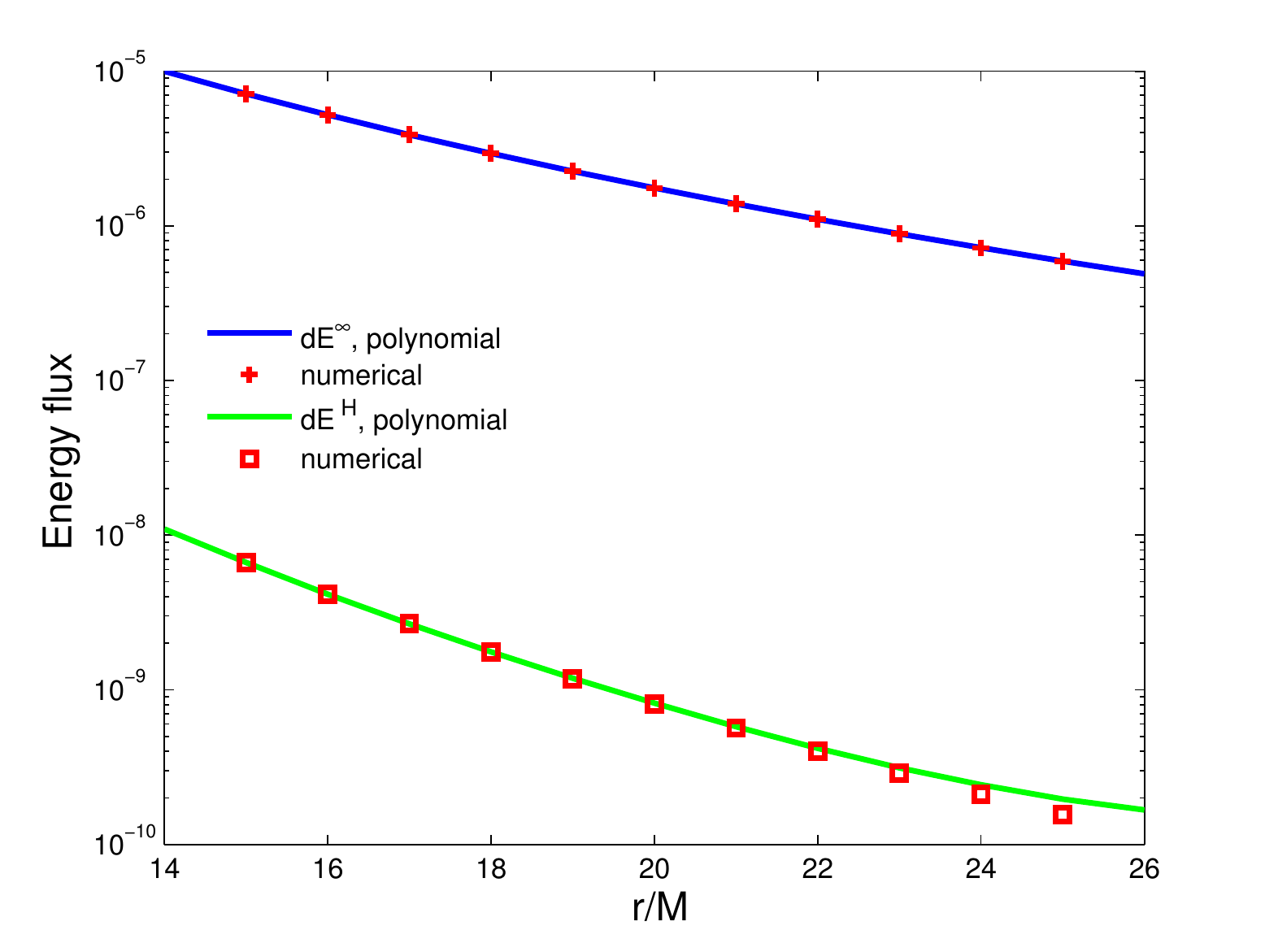}
\includegraphics[height=2.0in]{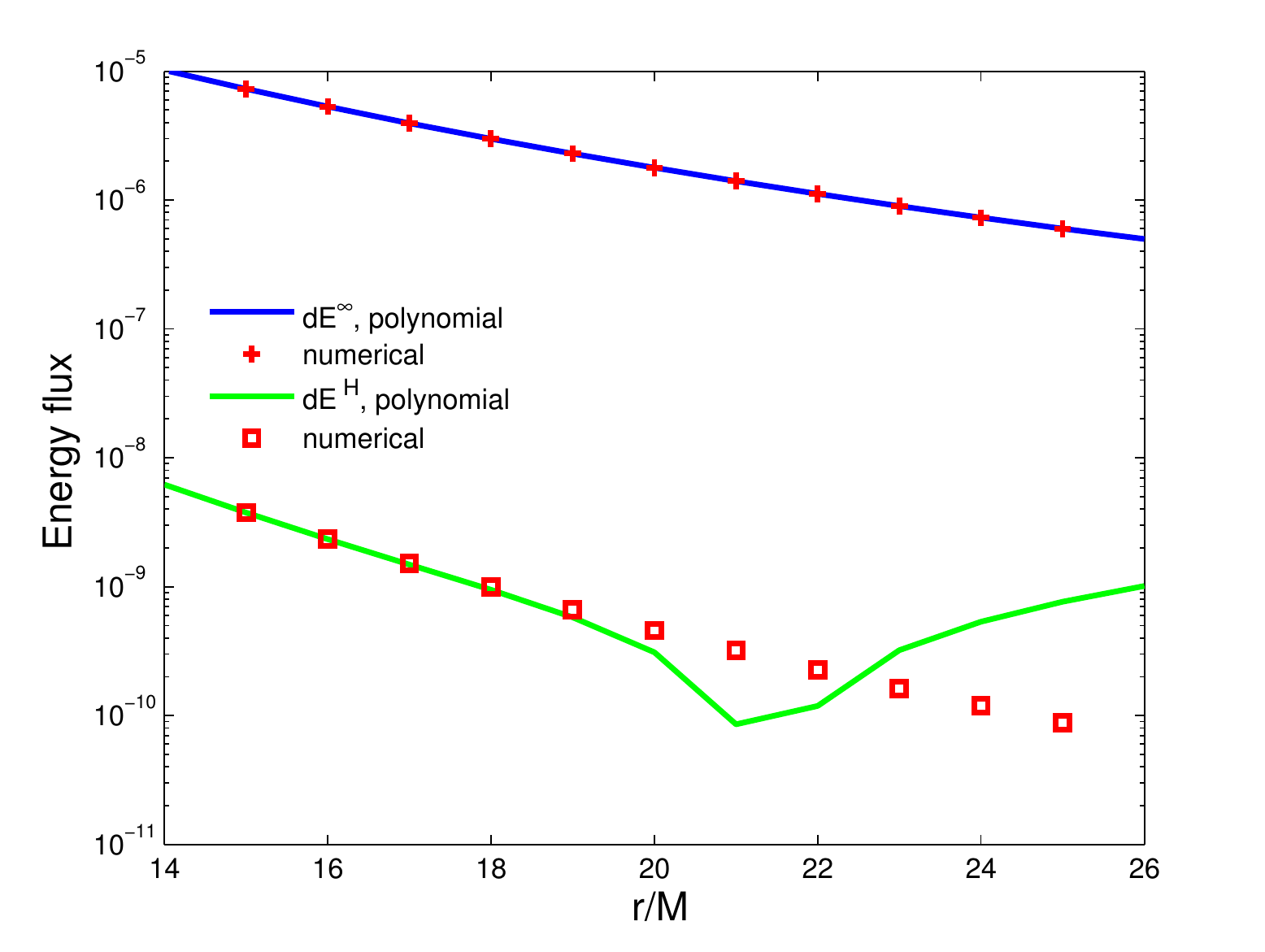}
\includegraphics[height=2.0in]{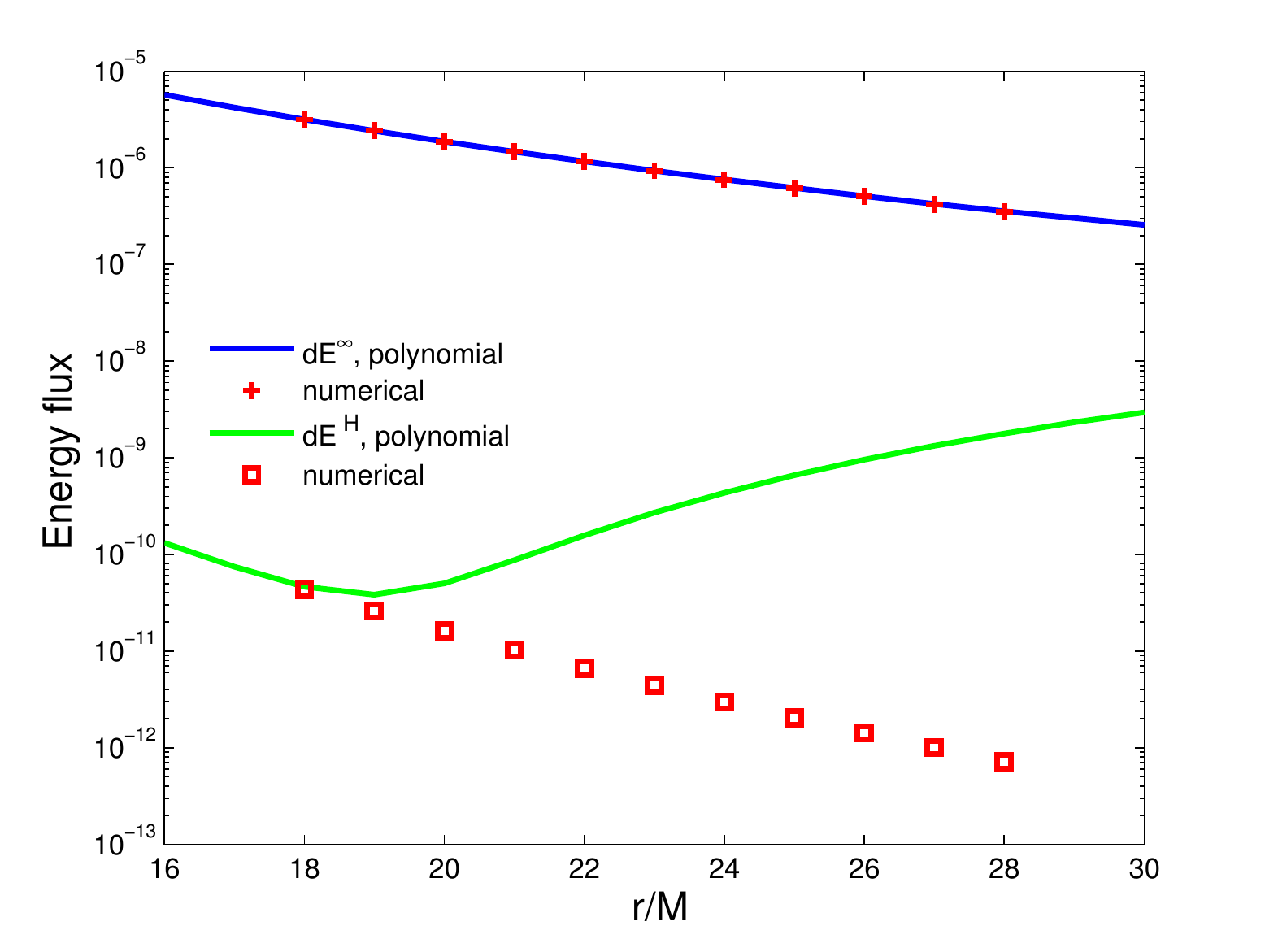}
\includegraphics[height=2.0in]{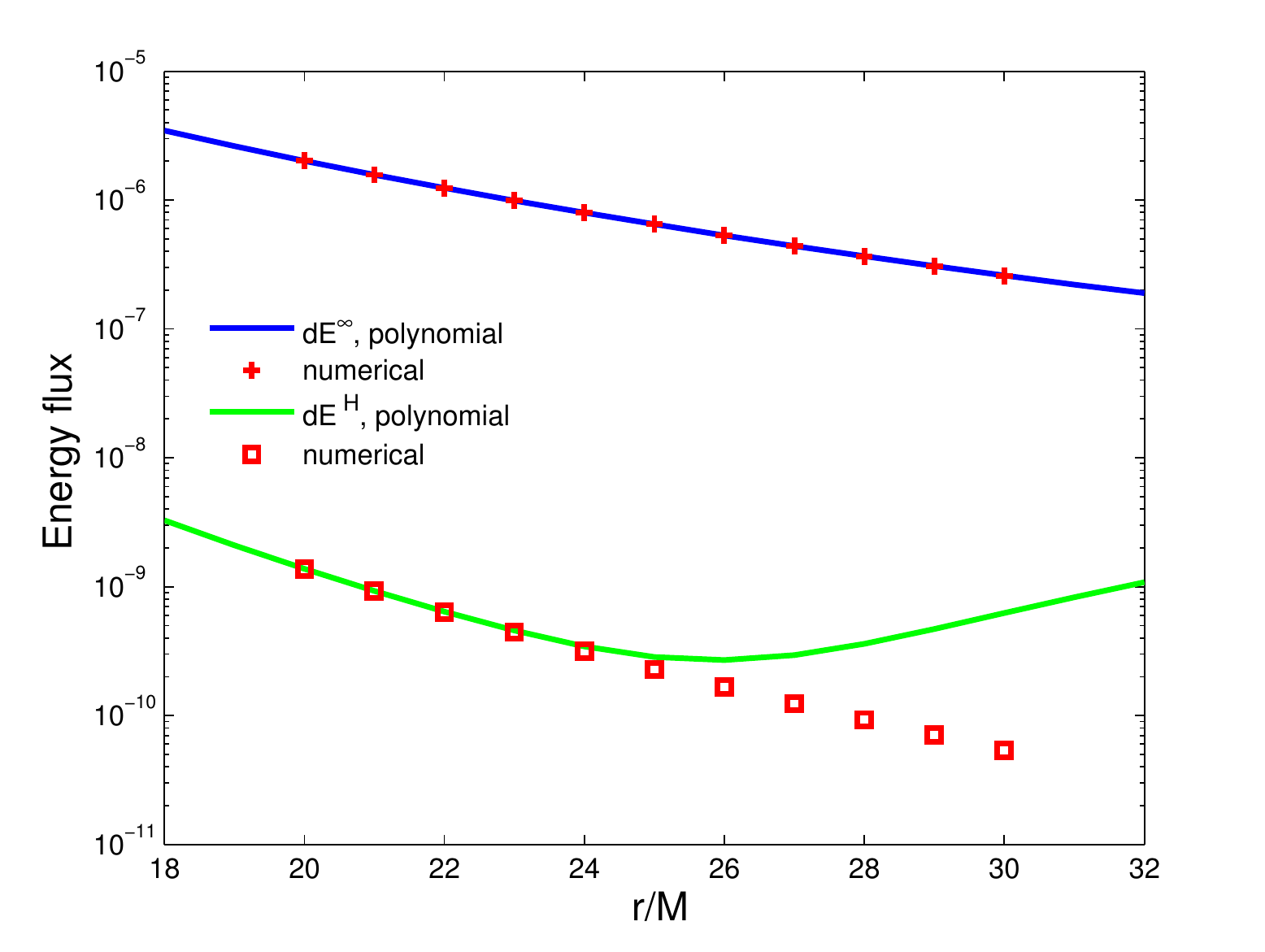}
\caption{Reliability of the polynomials listed in Tab. \ref{E8a9} and Tab. \ref{Eha9} when they are extendedly used to the further area from the central black hole.}  \label{extendfit}
\end{center}
\end{figure}

The polynomials above are obtained from the numerical data of 11 points between ISCO and $10 M+r_\text{ISCO}$. Actually one can choose any range which is interesting for the research to fit out the corresponding polynomials. One can also choose the number of points and the order of polynomials based on the accuracy requirement. However, the polynomials we obtained may be used directly to the neighbor area (except for  the area inside the ISCO). In Fig. \ref{extendfit}, we show the validity of our polynomials £¨fitted from the data between $r_{\rm ISCO}$ and $r_{\rm ISCO} + 10 M $) in the further area. We can see that the polynomials for $\dot{E}^\infty$ fitted from the numerical data can be used directly to the further area. At the same time, the polynomials for  the energy flux down to the horizon do not perform very well in the further area. However,  $\dot{E}^\text{H}$ is much less than $\dot{E}^\infty$ (only $10^{-5}$ of the latter). In many cases, one can just simply omit $\dot{E}^\text{H}$.

In addition, one can also fit the energy fluxes and waveforms by the polynomials with post-Newtonian parameter $x \equiv (v/c)^2 = (GM\Omega/c^3)^{2/3}$, then Eqs. (\ref{Epoly}) and (\ref{hpoly}) are transferred to
\begin{align} \label{Epolyx}
\dot{E}^{\infty}&=\sum_{i=0}^{n}a'_i x^i,\quad \dot{E}^{\text{H}}=\sum_{i=0}^{n}b'_i x^i , \\
\text{Re}[H_{lm}]&=\sum_{i=0}^{n} R'^{i}_{lm} x^i , \quad \text{Im}[H_{lm}]=\sum_{i=0}^{n} I'^{i}_{lm} x^i . \label{hpolyx}
\end{align}

The Eqs. (\ref{Epoly} - \ref{hpolyx}) are essentially post-Newtonian expansions. However, all coefficients are obtain from numerical fitting of the Teukolsky-based data to guarantee the accuracy, in contrast to the analytical expressions of the post-Newtonian approximation like as the factorized-resummation ones.

\comment{
\begin{figure}
\begin{center}
\includegraphics[height=2.0in]{e8_a9_nu0_r.pdf}
\includegraphics[height=2.0in]{de8_a9_nu0_r.pdf}
\caption{match total energy fluxes for $a=0.9$.} \label{a9fit}
\end{center}
\end{figure}
\begin{figure}
\begin{center}
\includegraphics[height=2.0in]{e8_a7_nu0_r.pdf}
\includegraphics[height=2.0in]{de8_a7_nu0_r.pdf}
\caption{match total energy fluxes for $a=0.7$.} \label{a7fit}
\end{center}
\end{figure}
\begin{figure}
\begin{center}
\includegraphics[height=2.0in]{e8_a0_nu0_r.pdf}
\includegraphics[height=2.0in]{de8_a0_nu0_r.pdf}
\caption{match total energy fluxes for $a=0$.} \label{a0fit}
\end{center}
\end{figure}
\begin{figure}
\begin{center}
\includegraphics[height=2.0in]{e8_a_9_nu0_r.pdf}
\includegraphics[height=2.0in]{de8_a_9_nu0_r.pdf}
\caption{match total energy fluxes for $a=-0.9$.} \label{a_9fit}
\end{center}
\end{figure}
}

\section{Orbital evolution and waveform}
During the inspiralling process, the orbit of small body is semi-circular, and the frequency-domain Teukolsky based waveform is highly accurate in this process. As discussed in the last section, we use the 10th polynomials to replace the original numerical Teukolsky fluxes and waveforms. For producing the 10th polynomials, firstly we need flux and waveform data of the 11 points during the evolution. These data are calculated by the Teukolsky equation. Once the evolution area is decided, the 11 interpolation points are generated by the Chebyshev nodes in this work.  Calculating the Teukolsky-based fluxes and waveforms at 11 points numerically only takes few seconds by a desktop.

Using the data on these 11 points, we can give out the 10th-order polynomials immediately. With the flux-polynomials at hand, we can use the EOB dynamics to evolve the orbits very fastly. The well-known EOB formalism was first introduced by Buonanno and
Damour more than ten years ago to model comparable-mass black hole
binaries \cite{Buonanno1,Buonanno2}, and was also applied in small
mass-ratio systems \cite{Nagar07,Bernuzzi10,Bernuzzi11a,Bernuzzi11b,Damour07,Damour09}.  The EOB dynamical evolution equations under radiation reaction for a quasi-circular orbit can
be given as \cite{BCD06,Panyi11b}
\begin{align}
\dot{r}&=\frac{\partial H_\text{EOB}}{\partial p_r},\label{rdot}\\
\dot{\phi}&=\frac{\partial H_\text{EOB}}{\partial p_\phi},\label{fdot}\\
\dot{p_r}&=-\frac{\partial H_\text{EOB}}{\partial r}+\mathcal{F}_\phi \frac{p_r}{p_\phi},,\label{prdot}\\
\dot{p_\phi}&=\mathcal{F}_\phi, \label{pfdot}
\end{align}
where $\mathcal{F}_\phi=\dot{E}/\dot{\phi}$, and $\dot{E}$ is the energy-flux of gravitational radiation.
For a non-spinning test particle, the Hamiltonian is
\begin{align}
H_{\text{NS}}=\beta^i p_i+\alpha\sqrt{\mu^2+\gamma^{ij}p_i
p_j},\label{HNS}
\end{align}
where $\mu = m_1 m_2/M$, $m_1, ~ m_2$ are the masses of two bodies respectively, $M = m_1+m_2$ and
\begin{align}
\alpha&=\frac{1}{\sqrt{-g^{tt}}},\\
\beta^i &=\frac{g^{ti}}{g^{tt}},\\
\gamma^{ij}&=g^{ij}-\frac{g^{ti}g^{tj}}{g^{tt}},
\end{align}
$g^{\mu\nu}$ is the inverse Kerr metric.

In our previous ET codes, the energy fluxes are obtained from Eq. (\ref{energyflux}) by calculating the Teukolsky equation numerically at every time step. Then, the fluxes are sourced to the EOB dynamical equations (\ref{rdot}-\ref{pfdot}) to drive the particle inspirals into the central black hole. This method will cost a lot of CPU time on the calculation of the Teukolsky equation. Now we use the flux-polynomials (\ref{Epoly}) or (\ref{Epolyx}) as the source in the EOB dynamical equation. For getting the coefficients of (\ref{Epoly}) or (\ref{Epolyx}), we need also to calculate the Teukolsky equation but only at a few points (11 points are chosen in this paper, calculation was finished in few seconds). With the
flux-polynomials at hand, the EOB dynamical equations then can be driven in a very fast way. At every time step, we calculate the waveforms by polynomials (\ref{hpoly}) or (\ref{hpolyx}). We list our numerical algorithm here for clarity:

\bn (1) determine the calculating area of the EMRIs by distance ($r_0,  r_\text{end}$) or GW frequency ($f_0^\text{GW}, f_\text{end}^\text{GW}$);

\bn (2) choose $n'$ points (Chebyshev nodes) in this area to calculate the Teukolsky-based energy fluxes and waveform, then to fit out $n$ order flux- and waveform- polynomials based on the accuracy requirement ($n' > n $).

\bn (3) calculate the dynamics of the small body by solving the EOB dynamical Eqs. (\ref{rdot})-(\ref{pfdot}) with the source-term obtained in (2) from the initial point ($r_0$, $\phi_0$, ${p_r}_0$, ~${p_\phi}_0$);

\bn (4) compute the waveforms by the polynomials obtain in (2) at every evolution temporal point until to the end.

\bn (5) if necessary, we need an iteration method for the correction of non-circular orbit: take the evolved data $\Omega_\phi$, $v_r$ at the $N'$ points instead of the original $\Omega^{\rm circ}_\phi, v^{\rm circ}_r = 0$
into the step (2) and repeat all the remainning steps.

\begin{figure}
\begin{center}
\includegraphics[height=2.0in]{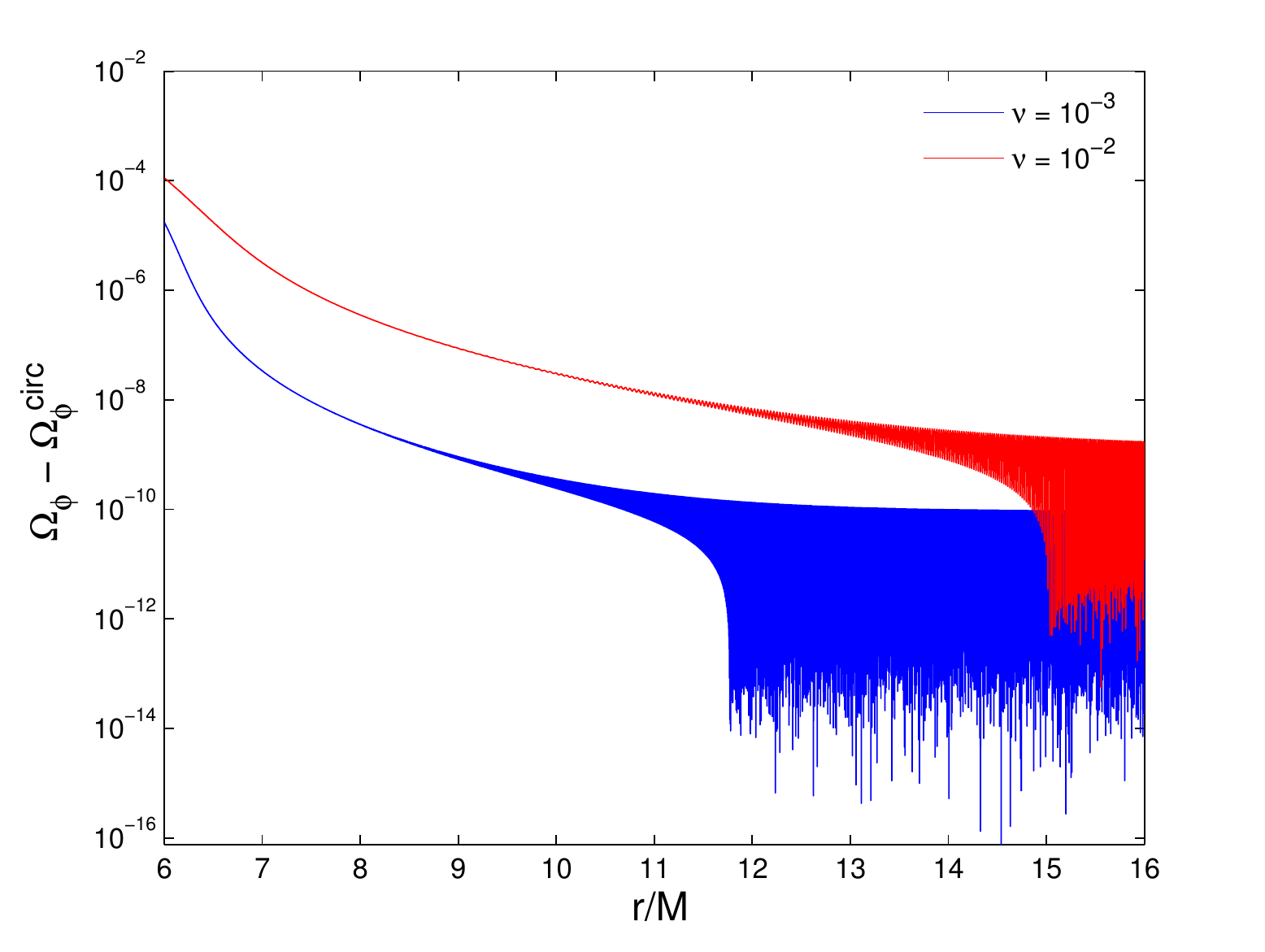}
\includegraphics[height=2.0in]{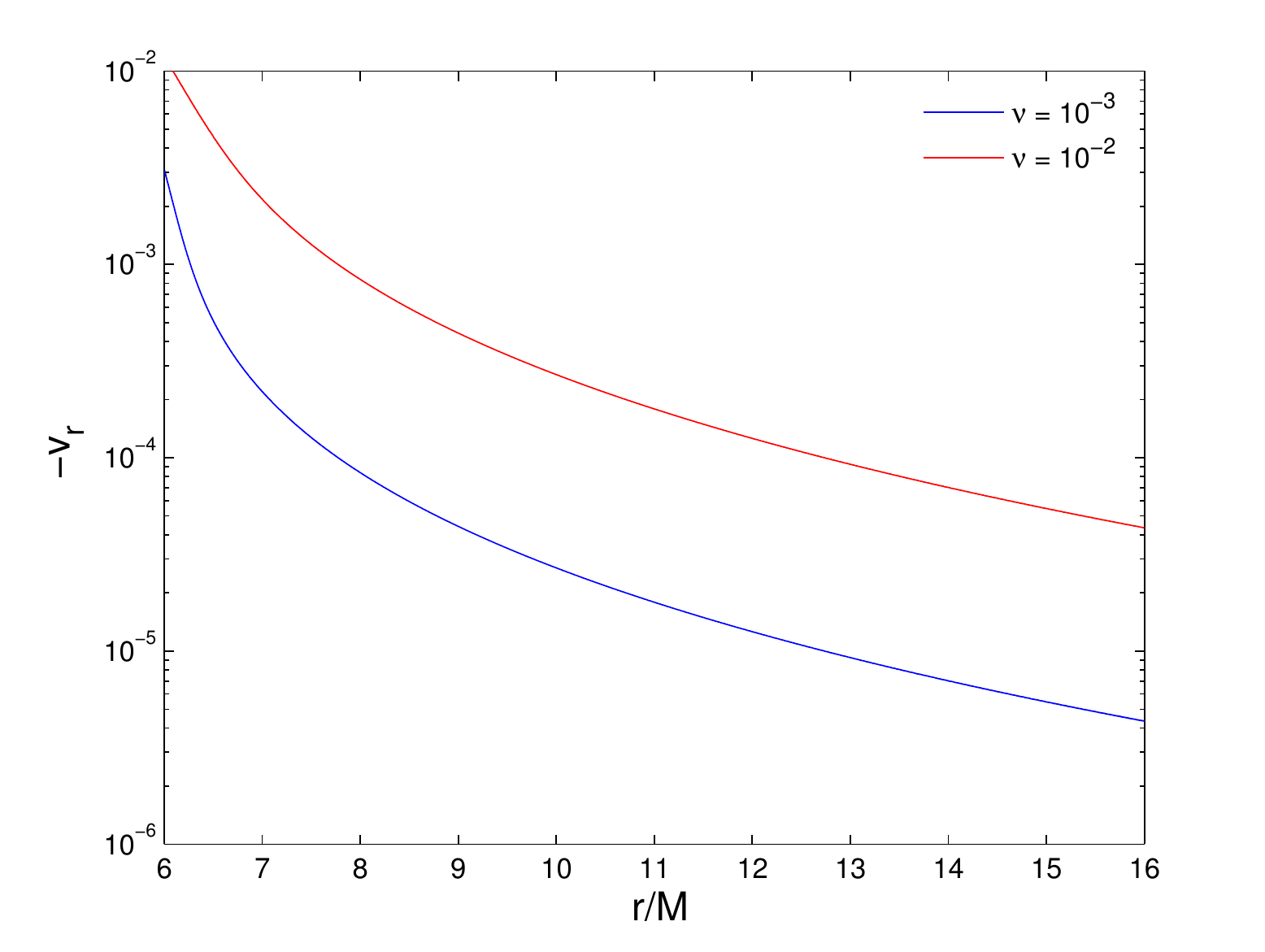}
\caption{The difference of the practical orbital frequency and the circular orbital one (the left panel); The radial velocity (the right panel).} \label{frequency}
\end{center}
\end{figure}

We use the orbital frequency $\Omega^{\rm circ}_\phi$ from circular orbital condition to calculate the Teukolsky equation. However, during the evolution process, the small body has radial velocity and the orbit is not exactly circular. Therefore the practical orbital frequency calculated by (\ref{fdot}) is different from $\Omega^{\rm circ}_\phi$. This is why we need an iteration procedure in the step (5). For mass-ratio $\mu / M \lesssim 10^{-3}$, the iteration is not necessary, but for the one of $10^{-2}$, we recommend to do the iteration. Please see Fig. \ref{frequency} for the comparison of the orbital frequency and radial velocity.

\begin{figure}
\begin{center}
\includegraphics[height=2.0in]{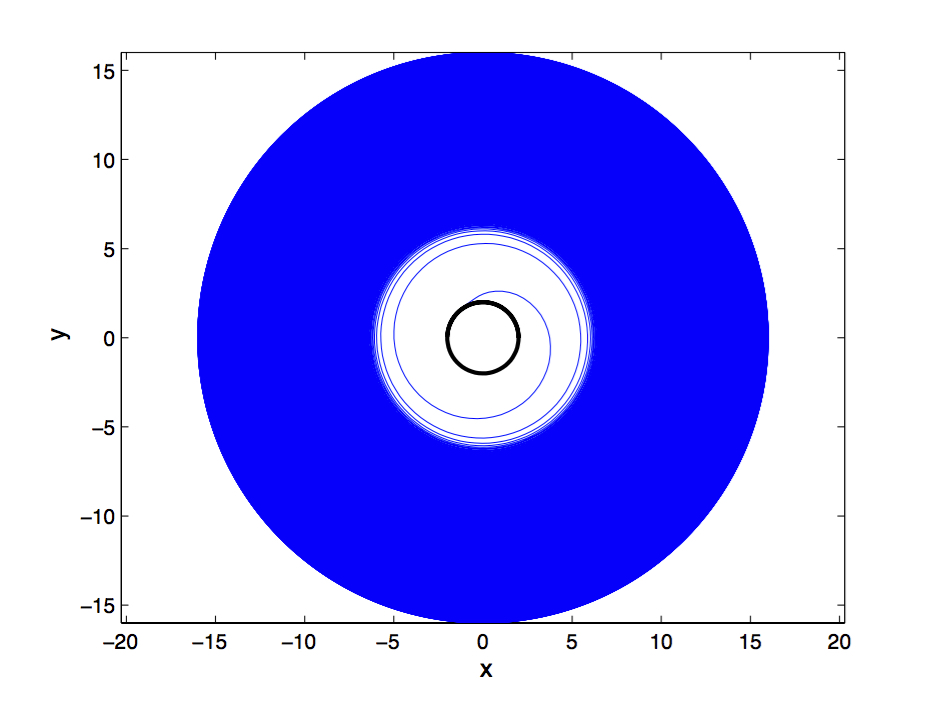}
\includegraphics[height=2.0in]{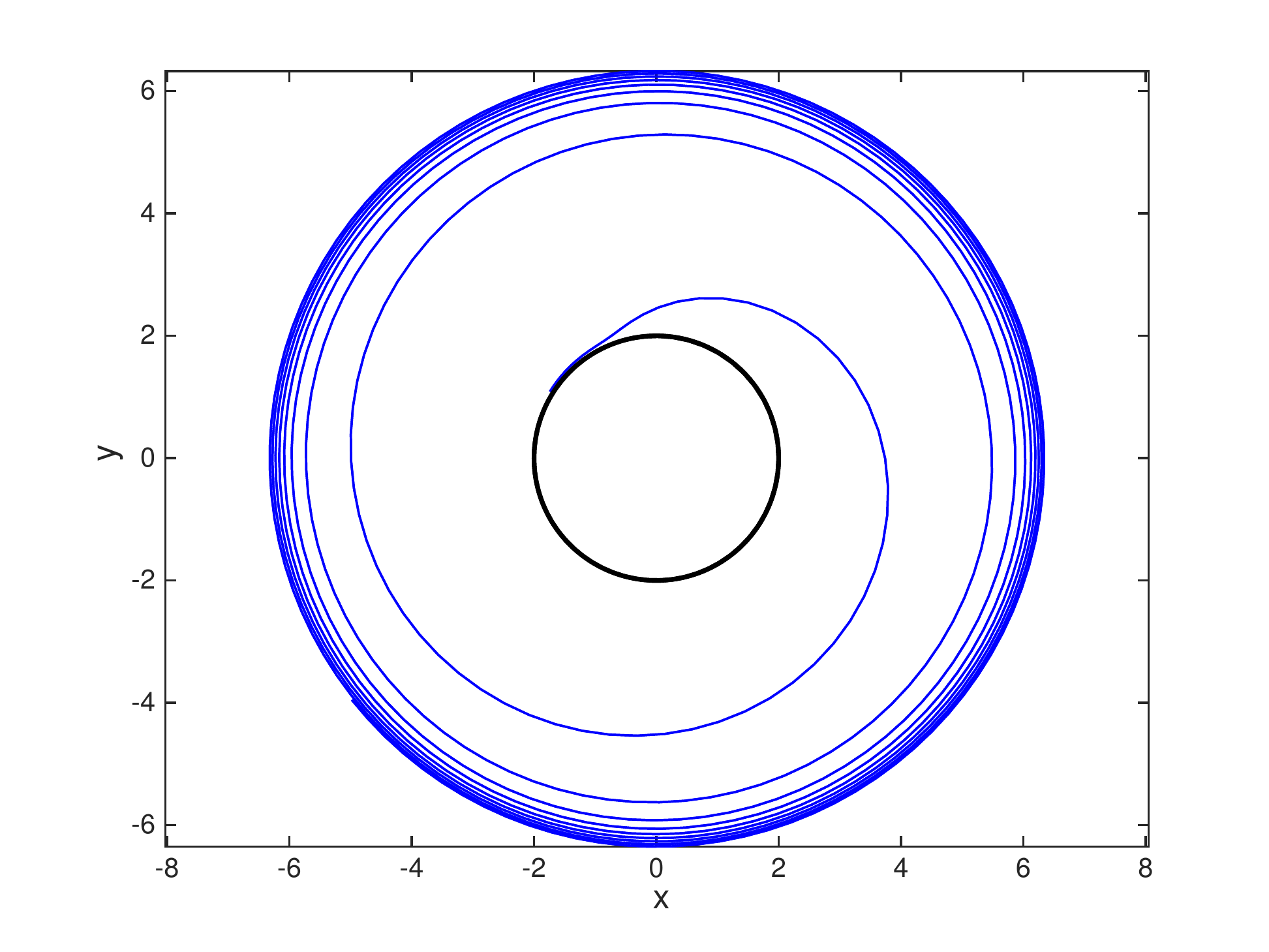}
\caption{Orbital evolution of the EMRIs (mass ratio 1/1000) for a = 0; the right panel shows the details of the orbit evolution at the final time.} \label{evolution}
\end{center}
\end{figure}

In Fig. \ref{evolution}, as an example, we show the orbital evolution of EMRIs with $\mu/M = 10^{-3}$ for $a = 0 $. Because the plunge process is dominated by the conservation dynamics, the plunge orbits passed the ISCO are archive here just by ignoring the radiation-reaction.


\begin{figure}
\begin{center}
\includegraphics[height=2.0in]{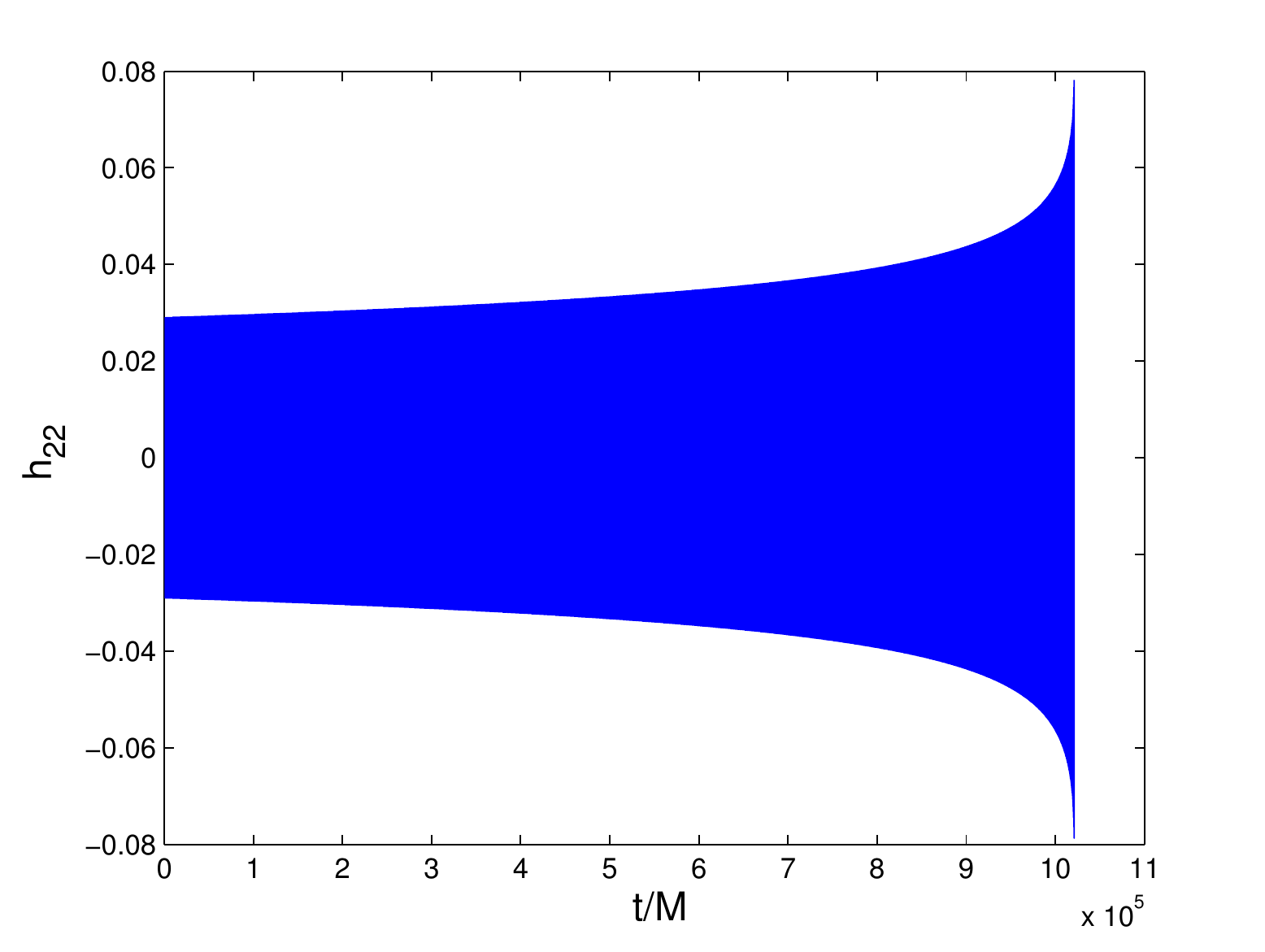}
\includegraphics[height=2.0in]{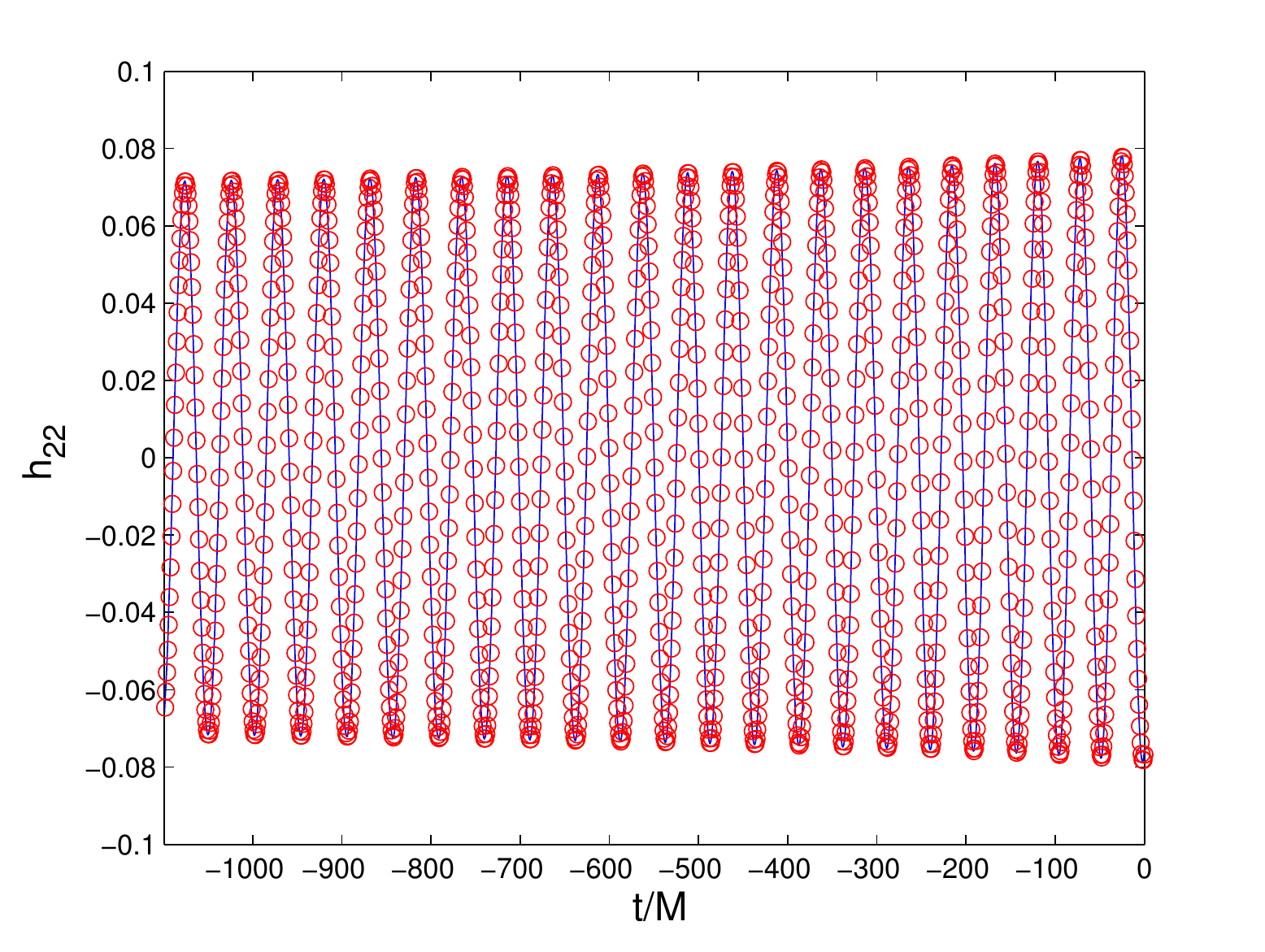}
\caption{Waveforms (left panel: plus-polarized part of $h_{22}$) of the orbital evolution in figure \ref{evolution} for a = 0; right panel: a part of the waveform --- the solid blue line is the polynomial waveforms, and the red circles represent the numerical Teukolsky ones. The time-coordinate $t = 0$ means the moment when the small body arrives at the ISCO.} \label{waveforms}
\end{center}
\end{figure}

The corresponding waveforms are demonstrated in Fig. \ref{waveforms}. All these evolutions are archive in 1 second by one CPU of a desktop. In the final year of the inspiral, an EMRI waveform has $10^5$ circles \cite{Gair04}. The computation time of such waveform is less than 300 seconds (single CPU) for an EMRI. However, the complexity of EMRI waveforms makes this procedure challenging. The inspiral waveform depends on 14 different parameters \cite{Glampedakis02}. Based on the analysis of \cite{Gair04}, if we assume that only about eight of these 14 parameters affect the phase evolution, it will need $300 \times (10^{5})^8$ s to produce all waveform templates with a single CPU! Only when we assume there are two to three parameters, this computation time becomes acceptable by using thousands of CPUs with parallel technology. If we use the standard Teukolsky techniques, it will take about a few tens of days to compute one year¡¯s evolution by using our codes with one CPU! It is much longer than the new procedure developed in this paper. In this sense, our new technique makes great progress in saving CPU time, though it is still far from practical demands.

To confirm the validation of our polynomial waveforms, we compare the fitting polynomials with the numerical Teukolsky waveform near the ISCO. We find that for the mass-ratio 1:1000, the match of our model with the numerical waveforms looks quite good (see the left panel of figure \ref{waveforms}). For confirmation, based on the matched-filter technology (an optimal method when searching for known signals in noisy data), we use the spectral noise density of LISA \cite{SNR} to calculate the overlap between our polynomial waveforms and the direct Teukolsky ones.  According to the basic set-up in matched filtering, the overlap $O_{a,b}$ between ?two time series of signals a and b is defined as,
\begin{align}
O_{a,b} = \frac{(a|b)}{\sqrt{(a|a)(b|b)}} \,,
\end{align}
where the product $(a|b)$ is given as
\begin{align}
(a|b) = 2 \int ^\infty_0 {\frac{\tilde a(f) \tilde b^*(f)+\tilde a^*(f) \tilde b(f)}{S_n (f)} df} \,.
\end{align}
The overhead tildes stand for the Fourier transform and the star stands for a complex conjugation. The quantity Sn(f) is the spectral noise density curve taken from [32]. From these equations, the overlaps of the two kinds of waveforms (our waveform polynomials and the Teukolsky one) are $99.95\%, ~97.01\%, ~99.76\%$ and $100\%$ for $a = 0.9, ~0.7,~ 0$ and $-0.9$ respectively.
This means that our polynomial waveforms are faithful during the whole evolution process. This also confirm our previous claim: for the mass-ratio  $ \lesssim 10^{-3}$, we need not an iteration to correct the quasi-circular approximation. The results of comparison are plotted in Fig. \ref{waveformcomp}. However,  as we have claimed, for mass-ratio around 1:100, we suggest that one should take the step (5) to obtain the more accurate waveforms because of the large radial velocity.

\section{Conclusions and discussions}
In this paper, we use the Teukolsky-based fluxes and waveforms at a few points on the evolution route of the EMRI to fit out a set of polynomials for fluxes and waveforms. A circular orbital condition is adopted to solve the orbital parameters for numerical calculation of the Teukolsky equation in a frequency domain.

Essentially, these flux and waveform polynomials are also a kind of PN expansion but all coefficients are obtained from the fitting of numerical data. Usually the PN coefficients are calculated from analytical expressions, such as the resummation PN waveform and Fujita¡¯s 11th PN results for EMRIs \cite{Fujita12}. As we can see from Fig. \ref{E8fit}, our 10th or 12th-order polynomials are much better than the resummation PN fluxes especially for the spin black hole cases. Comparing the results shown in Fig. 4 of \cite{Fujita12}, we can find  that our results are also more accurate than the 11th PN analytical fluxes which have very long expressions.

Yunes et al used the numerical Teukolsky-based fluxes to calibrate higher-order PN coefficients and add them to resummed PN analytical expressions \cite{Yunes10, Yunes11}. Though their results are quite good, the accuracy of our flux polynomials is a little better than theirs. Furthermore, the fitting of the coefficients of polynomials costs less CPU time than their calibration method. Solving the Teukolsky equation to give a few sets of fluxes and wave- forms by the semi-analytical method introduced in section 2 just needs a few seconds, and the fitting process almost does not add extra CPU time. However, Yunes has mentioned that they need $O(10)$ minutes to complete calibration \cite{Yunes11}.

Just for demonstration, in the present paper we use a total of 11 points to fit out a set of 10th-order polynomials for fluxes and waveforms. However, using more points can fit out more accurate polynomials. For examples, 10th or 12th-order polynomials obtained from 20 points will give out better fluxes and waveforms. This will only add a little CPU time (twice of the case of 11 points).
With these polynomials at hand, EMRIs can be evolved in a very fast way and the waveforms can be extracted at the same time. The parameters listed in tables (\ref{E8a9}-\ref{RH22}) can be used directly for the GW data analysis. In the present paper, we only fit out the one- dimensional polynomials of r. In principle, one can try to fit out two-dimensional polynomials of both r and a. Unfortunately we find that the coefficients of some polynomials vary suddenly while a just changes a little. We will leave this to future work. For many different a of Kerr black holes, principally we can build a database of flux and waveform polynomials by scanning a large number of different Kerr parameters (each a only takes a few seconds to produce the polynomials). Our fast and accurate polynomial model could be useful in the waveform-template calculations for future GW detectors such as eLISA, Taiji \cite{taiji} and Tianqin \cite{tianqin}.
\section*{Acknowledgments}
This work was supported by the NSFC (No. U1431120, No.11273045).

\end{document}